\documentclass[
	aps, prd, reprint,
	10pt, notitlepage, a4paper,
         floats, floatfix,
	amsmath, amssymb, amsfonts, eqsecnum,
	superscriptaddress,
	showpacs, showkeys,
	nofootinbib,
 	longbibliography,
]{revtex4-1}

\usepackage{aas_macros, float}
\usepackage[usenames,dvipsnames]{xcolor}
\usepackage{xspace} 
\usepackage{caption}
\usepackage{subcaption}
\usepackage{bm,graphicx} 
\usepackage[utf8]{inputenc} 

\xdefinecolor{mylinkcolor}{rgb}{0,0,0.5}
\usepackage[
	bookmarksnumbered, bookmarksopen, bookmarksopenlevel=2,
	breaklinks=true,
	colorlinks=true, filecolor=mylinkcolor, citecolor=mylinkcolor,
	linkcolor=mylinkcolor, urlcolor=mylinkcolor, menucolor=mylinkcolor,
]{hyperref}

\def\ee{\end{equation}}
\def\eea{\end{eqnarray}}
\def\be{\begin{equation}}
\def\bea{\begin{eqnarray}}

\allowdisplaybreaks

\begin{document}

\newcommand{\AEI}{\affiliation{Max-Planck-Institute for Gravitational Physics (Albert-Einstein-Institute),
\\ Am M{\"u}hlenberg 1, 14476 Potsdam-Golm, Germany, EU}}

\newcommand{\GRAPPA}{\affiliation{GRAPPA Institute of High-Energy Physics, University of Amsterdam, Science Park 904, 1098 XH Amsterdam, The Netherlands, EU}}

\newcommand{\DITF}{\affiliation{Delta Institute for Theoretical Physics, Science Park 904, 1090 GL Amsterdam, The Netherlands, EU}}

\newcommand{\ITF}{\affiliation{Institute for Theoretical Physics, Utrecht University, Princetonplein 5, 3584 CC Utrecht, The Netherlands, EU}}

\newcommand{\UU}{\affiliation{Department of Physics, Utrecht University, Princetonplein 1, 3584 CC Utrecht, The Netherlands, EU}}

\newcommand{\GRASP}{\affiliation{Institute for Gravitational and Subatomic Physics (GRASP),\\ Department of Physics, Utrecht University, Princetonplein 1, 3584 CC Utrecht, The Netherlands, EU}}

\newcommand{\Nikhef}{\affiliation{Nikhef, Science Park, 1098 XG Amsterdam, The Netherlands, EU}}

\title{Effect of dynamical gravitomagnetic tides on measurability of tidal parameters for binary neutron stars using gravitational waves}

\date{\today }

\author{Pawan Kumar Gupta}
\email{p.gupta@nikhef.nl}
\Nikhef \GRASP

\author{Jan Steinhoff}
\email{jan.steinhoff@aei.mpg.de}
\homepage{http://jan-steinhoff.de/physics/}
\AEI

\author{Tanja Hinderer}
\email{t.p.hinderer@uu.nl}
\ITF  

\begin{abstract}
 Gravitational waves (GWs) from binary neutron stars (NSs) have opened unique opportunities to constrain the nuclear equation of state by measuring tidal effects associated with the excitation of characteristic modes of the NSs. This includes gravitomagnetic modes associated with the Coriolis effect, whose frequencies are proportional to the NS's spin frequency, and for which the spin orientation determines the subclass of modes that are predominantly excited. We advance the GW models for these effects that are needed for data analysis by first developing a description for the adiabatic signatures from gravitomagnetic modes in slowly rotating NSs. We show that they can be encapsulated in an effective Love number which differs before and after a mode resonance. Combining this with a known generic model for abrupt changes in the GWs at the mode resonance and a point-mass baseline leads to an efficient description which we use to perform case studies of the impacts of gravitomagnetic effects for measurements with Cosmic Explorer, an envisioned next-generation GW detector. We quantify the extent to which neglecting (including) the effect of gravitomagnetic modes induces biases (significantly reduces statistical errors) in the measured tidal deformability parameters, which depend on the equation of state. Our results substantiate the importance of dynamical gravitomagnetic tidal effects for measurements with third generation detectors.

\end{abstract}

\maketitle


\section{Introduction}

The gravitational wave (GW) discovery of the binary neutron star (NS) inspiral GW170817~\cite{LIGOScientific:2017vwq} provided, for the first time, a purely gravitational channel for probing the properties of dense matter in NS interiors, whose equation of state remains poorly constrained~\cite{Geesaman:2015fha,Nupecc}. While this event provided the first empirical constraints with GWs, more precise measurements of the equation of state will become possible as existing detectors (such as LIGO~\cite{TheLIGOScientific:2014jea}, Virgo~\cite{TheVirgo:2014hva},KAGRA~\cite{Aso:2013eba}) improve in sensitivity in the coming years~\cite{KAGRA:2013rdx} and next-decade's envisioned third generation facilities such as Einstein Telescope~\cite{Punturo:2010zza} and Cosmic Explorer~\cite{Reitze:2019iox} become operational. These next-generation detectors will have a much higher sensitivity and wider bandwidth, which will open opportunities for transformative insights into dense matter under extreme gravity~\cite{Maggiore:2019uih,Sathyaprakash:2019yqt,Kalogera:2021bya}.  Realizing this science potential critically relies on advancing theoretical models of the GWs from binary systems with matter effects, which are needed to extract information about the source properties from the data, as reviewed in~\cite{LIGOScientific:2019hgc}. To date, GW measurements have only been sensitive to the dominant effects of NS matter on the signals and had relatively large statistical errors, causing systematic errors due to shortcomings in the modeling to be subdominant~\cite{LIGOScientific:2018mvr}. However, similar measurements at a higher sensitivity or with future detectors will require models that are significantly more accurate and include more realistic physics to enable more stringent constraints on NS matter and avoid biases in the interpretation. 

During a binary inspiral, the GW signatures of the properties of matter in NSs are due to spin and tidal effects. Tidal effects encompass various phenomena associated with the resonant or non-resonant excitation of characteristic oscillation modes of the NS, whose properties in turn depend on the properties of dense subatomic matter. The modes are driven by the tidal fields of the companion, which vary in time due to the orbital motion and can be decomposed into gravito\emph{electric} and -\emph{magnetic} fields depending on their parity properties. The former are involved in the dominant tidal effects due to the fundamental modes of the NS, which have the strongest tidal couplings and relatively high resonance frequencies that leave their excitation non-resonant for most of a quasi-circular inspiral~\cite{Flanagan:2007ix,Steinhoff:2016rfi,Pratten:2021pro}. By contrast, gravitomagnetic tidal fields associated with relativistic frame-dragging effects lead to the excitation of 
inertial modes of NSs whose frequencies are proportional to the spin~\cite{Kokkotas:1999bd, Andersson:2000mf,Idrisy:2014qca,Lee:2002fp,Kokkotas:2015gea} and will thus invariably pass through resonances in binary inspirals. The resonant energy and angular momentum transfer between the modes, orbit, and GWs leads to comparatively sudden changes in the GW frequency evolution, thus contributing a small but distinctive feature to the signals. 

There has been much previous work on gravitomagnetic modes of NSs, which are associated with the Coriolis effect and include inertial modes such as the 'r-modes'~\cite{1981A&A....94..126P,Ho:1998hq,Schenk:2001zm,Lockitch:1998nq}. Racine and Flanagan~\cite{Flanagan:2006sb} computed the direct effects of the resonance on the dynamics and developed an effective waveform model for the resulting GW imprints. This model was recently revisited to assess the impact for measuring tidal deformabilities with next-generation detectors~\cite{Ma:2020oni, Poisson:2020eki, Poisson:2020mdi};
see~\cite{Yu:2017cxe} for use of the model for other classes of modes and ~\cite{Miravet-Tenes:2023kte} for studies of inertial modes in postmerger GWs. Studies have also modeled and examined the effect of nonresonant gravitomagnetic tides on the inferred tidal deformability~\cite{JimenezForteza:2018rwr}, and included them in an effective one body model~\cite{Akcay:2018yyh}. However, the conclusions were limited due to an interesting feature of the response of a NS's matter and spacetime to a gravitomagnetic tidal perturbation, which leads to two possible kinds gravitomagnetic tidal deformabilities depending on the assumptions on the state of the perturbed fluid~\cite{Damour:2009vw, Binnington:2009bb, Landry:2015cva, Landry:2015snx, Poisson:2016wtv, Pani:2018inf}. It turns out that the significance of these two tidal deformabilities is that both are relevant but in different ways for the asymptotic limits of the response before and after a gravitomagnetic mode resonance~\cite{Gupta:2020lnv}.

In this paper, we first derive an explicit expression for the effective gravitomagnetic response function characterizing the ratio of the induced current quadrupole moment to the gravitoelectric tidal field in the context of a binary system at large separation with arbitrary spin orientations and low spin magnitudes. The asymptotic limits of this response before and after resonance yield the relevant combinations of the gravitomagnetic tidal deformabilities in the different regimes. A new aspect in this paper is that we include these effects together with the direct resonance-induced changes in the GWs from~\cite{Flanagan:2006sb,Ma:2020oni}. A further difference compared to this previous work is that we map all EoS-dependent parameters that appear in the resonance expressions and were thus far only considered for Newtonian descriptions of NSs to their fully relativistic counterparts, which we employ for further studies of the impact of gravitomagnetic tides on parameter estimation. In general, multiple quadrupolar gravitomagnetic modes with azimuthal number $|m|=1, 2$ are resonantly excited in an inspiral, however, certain spin orientations mainly favor the excitation of only one of them, which can be exploited to simplify an initial exploratory study~\cite{Flanagan:2006sb}. We also make use of previous findings that for NSs, the equation of state information contained in gravitomagnetic Love numbers can be approximately related to the main tidal deformability $\Lambda$, which reduces the number of signal parameters~\cite{Yagi:2013sva,JimenezForteza:2018rwr}. Furthermore, as the full parameter estimation in the entire parameter space for binary NS signals in third-generation detectors is prohibitive and the largest constraints will come from events with a high signal-to-noise ratio, we use restricted Fisher matrix computations as a proxy for the statistical errors. While all of these assumptions are restrictive, the aim of our work is to scope out the importance of gravitomagnetic modes on GW measurements with third-generation detectors using a more realistic model of these effects than in previous such studies. 
We first estimate the plausible changes in the width of the posterior distributions when using full Markov Chain Monte Carlo (MCMC) pipelines versus the Fisher matrix, perform a number of sanity checks on the results, and compare with previous work.  We then study the impact of different mode resonances as well as the asymptotic adiabatic contributions on the accuracy with which tidal deformability can be measured in a few different case studies, and the biases incurred when neglecting the gravitomagnetic effects.

The paper is organized as follows. In Sec.~\ref{sec:Love number} we obtain the effective Love number, discuss its features, and the description of gravitomagnetic tidal effects far from resonance. In Sec.~\ref{sec:waveform} we incorporate these results into a frequency-domain waveform model. We discuss the data analysis framework in Sec.~\ref{sec:gaussian} and the results in Sec.~\ref{sec:results}. Section~\ref{sec:summary} contains our conclusions and outlook. 

Unless otherwise specified, we use geometric units $G=c=1$. We use capital Latin letters from the middle of the alphabet $I, J, K, \ldots$ to denote spatial components of a tensor expressed in the rest frame of a NS. These indices are raised and lowered with the flat Cartesian three-metric $\delta_{IJ}$, thus, their up or down placement has no meaning. We use the Einstein summation convention that repeated indices are implied to be summed over. We also use round brackets around indices to denote their symmetrization, for instance, for two vectors $x^I$ and $v^J$ we denote $x^{(I}v^{J)}=(x^I v^J+x^J v^I)/2$.

\section{Effective Gravitomagnetic Love number}
\label{sec:Love number}
In this section, we review the identification of an effective magnetic Love number based on a fully relativistic formalism for slowly rotating bodies~\cite{Gupta:2020lnv} and calculate an explicit expression for the case of the leading-order gravitomagnetic tidal fields in a binary system. We also derive the adiabatic limits of these results for arbitrary spin orientations using an orbit-averaging procedure. Our results are based on considerations to linear order in the spins and focus on the quadrupole which is expected to give the largest effect. The entire framework we use is adapted to approximations based on the hierarchy of length- and timescales during the early part of a binary inspiral, see~\cite{Gupta:2020lnv} for more details.

\subsection{Definition of gravitomagnetic Love numbers}
A NS immersed in an external gravitomagnetic tidal field $B_{IJ}$ will develop an induced flux quadrupole moment $\mathcal{J}_{IJ}$. The  gravitomagnetic quadrupolar Love number, which we denote by $\sigma$, is defined as the ratio
\begin{subequations}
\label{eq:defLove}
\begin{equation}
\sigma=\frac{1}{2}\frac{ \mathcal{J}^{IJ} }{ B^{IJ}}. 
\end{equation}
Alternatively, $\sigma$ can be identifies as the coupling coefficient in the Lagrangian~\cite{Gupta:2020lnv} describing gravitomagnetic tides in the adiabatic limit according to the conventions
\begin{equation} 
  L_\text{ad}^{\cal B} = \frac{2 \sigma}{3} B_{IJ} B^{IJ} .
  \end{equation}
  \end{subequations}
Calculations of magnetic Love numbers based on relativistic perturbations of a NS revealed that magnetic quadrupolar Love numbers $\sigma$ can be of two types, depending on assumptions on the perturbed fluid inside the non-rotating neutron star~\cite{Damour:2009vw, Binnington:2009bb, Landry:2015cva, Landry:2015snx, Poisson:2016wtv, Pani:2018inf}. Restricting the fluid to remain static under perturbations leads to the static Love number $\sigma_\text{stat}$, while allowing it to be irrotational yields a different result $\sigma_\text{irrot}$. As discussed in~\cite{Gupta:2020lnv} and detailed below, both Love numbers are relevant for characterizing the gravitomagnetic tidal response of a NS asymptotically far from a mode resonance.

\subsubsection{Effective frequency-dependent Love number}
When going beyond the restriction to adiabatic limits, the tidal deformability generalizes to an effective frequency-dependent response function. Its particular form is given by considering the dynamics of the matter contributions to the flux quadrupole moment $Q_{\cal B}^{IJ}$ described by the Lagrangian given in Eq.(3.17) of~\cite{Gupta:2020lnv} as
\begin{align}\label{Lmag}
L^{{\cal B}} \approx - \frac{3}{32 (\sigma_{\rm irrot}-\sigma_{\rm stat})}\left[ \dot Q_{{\cal B}}^{I J} \dot{Q}_{{\cal B}}^{I J} - 2 \hat{\omega}^{\cal B} \Omega^{JK}\dot{Q}_{{\cal B}}^{I J}Q_{{\cal B}}^{KI}\right]
\nonumber \\ -\frac{1}{2} B_{IJ} \dot{Q}_{{\cal B}}^{I J}
+ \frac{2\sigma_{\rm stat}}{3} B_{IJ} B_{IJ} . 
\end{align}
Here, overdots denote proper time derivatives and the tensor $\Omega_{IJ}$ is related to the NS's spin frequency ${\bm \Omega}$ by
\begin{equation}
    \Omega_{IJ}=\epsilon_{IJK}\Omega^K,
\end{equation} where $\epsilon_{IJK}$ is the Levi-Civita permutation tensor. 
The dimensionless frequency quantity $\hat\omega_\mathcal{B}$ is given in terms of the quadrupolar gravitomagnetic mode frequencies in the co-rotating frame $\omega_{2 m}^\mathcal{B}$, where $l=2$ denotes the quadrupolar modes and $m$ the azimuthal mode number, by
\begin{equation}
\label{eq:omegahatdef}
  \hat\omega^\mathcal{B}=\frac{\omega_{2 m}^\mathcal{B}}{m\Omega}.
\end{equation}
In the Newtonian limit, the mode frequencies $\omega_{2m}^\mathcal{B}$ reduce to $\omega_{2m}^{\rm Newt}=-m\Omega/3$ in this frame, making \eqref{eq:omegahatdef} independent of $m$.

To obtain an effective adiabatic Lagrangian in the form of~\eqref{eq:defLove} we integrate the first term in~\eqref{Lmag} by parts and neglect the total time derivative. We then eliminate the acceleration $\ddot {Q}_{{\cal B}}^{I J}$ by using the oscillator equations of motion
\begin{equation}\label{eq:relEOM}
   \ddot{Q}_{{\cal B}}^{I J} -2 \hat{\omega}^{\cal B}\Omega^{K(I} \dot{Q}_{{\cal B}}^{J)K} = \frac{8 }{3}(\sigma_\text{irrot} - \sigma_\text{stat}) \dot B_{IJ}. 
\end{equation}
Substituting these equations of motion~\eqref{eq:relEOM} for $\ddot{Q}_{{\cal B}}^{I J}$ into the Lagrangian and omitting total derivatives leads to 
\begin{equation}
\label{eq:Lagronshell}
\bar L^{{\cal B}} \approx - \frac{1}{4} B_{IJ} \dot{Q}_{{\cal B}}^{I J}
+ \frac{2\sigma_{\rm stat}}{3} B_{IJ} B_{IJ} ,
\end{equation}
which is only valid for configurations of the system that satisfy the equations of motion~\eqref{eq:relEOM}. 

We identify the effective response function by requiring that the Lagrangian~\eqref{eq:Lagronshell} take the form of the adiabatic Lagrangian~\eqref{eq:defLove} with $\sigma$ replaced by an effective Love number
\begin{equation}
  \bar L^{{\cal B}} \stackrel{!}{=}  \frac{2 \sigma_\text{eff}}{3} B_{IJ} B_{IJ} .
\end{equation}
Omitting total derivatives, this leads to the identification of an instantaneous (inst) effective Love number
\begin{equation}
\label{eq:sigmaeffgen}
    \sigma_\text{eff}^{\rm inst}=  \frac{-\frac{3}{8}B_{IJ} \dot{Q}_{{\cal B}}^{I J}
+ \sigma_{\rm stat} B_{IJ} B_{IJ}}{B_{KL} B_{KL}}  .
\end{equation}
This result for an effective Love number still has undesirable features, for instance, it varies over an orbit and the definition is not unique due to the different ways of assigning the time derivatives up to total derivative terms. For example, the first term in the numerator of~\eqref{eq:sigmaeffgen} could equivalently be written as $3 \dot{B}_{IJ} Q_{{\cal B}}^{I J}/8$. 
These subtleties disappear when we impose that the above definitions hold only at the level of the orbit-averaged Lagrangians. Denoting the orbit-average by angular brackets, we 

define the effective Love number by
\begin{equation}
    \label{eq:effectivelovenumber}
    \sigma_\text{eff} = \sigma_{\rm stat}  -\frac{3}{8}\frac{\langle B_{IJ} \dot{Q}_{{\cal B}}^{I J}
\rangle}{\langle B_{IJ} B_{IJ}\rangle},
\end{equation}
with $\dot{Q}_{{\cal B}}^{I J}$ a solution to the equations of motion~\eqref{eq:relEOM}. The above definition of the effective Love number~\eqref{eq:effectivelovenumber} becomes more transparent when expressed in terms of the flux quadrupole defined in~\eqref{eq:defLove} which, as discussed in~\cite{Gupta:2020lnv}, is given by $\mathcal{J}^{IJ} = 2\sigma_{\rm stat} B_{IJ}- 3 \dot Q_{\cal B}^{IJ}/4 $. With this, 
\begin{equation}
\sigma_\text{eff} =\frac{1}{2} \frac{ \langle B_{IJ} \mathcal{J}^{IJ} \rangle}{\langle B_{IJ} B_{IJ}\rangle}, 
\end{equation}
which is directly analogous to the definition in the gravitoelectric case.

\subsection{Application to a binary system}

To obtain an explicit expression for the effective Love number requires specifying the relevant tidal field $B_{IJ}$. Here, we consider a binary system composed of the NS with mass $M_1$ and a point-mass companion $M_2$ at large orbital separation. We work in the center of mass frame of the NS and introduce a coordinate system in which the position of the center of mass of the companion is ${\bm z}(t)$ and its velocity is $\dot {\bm z}(t)$. The gravitomagnetic tidal field $B_{ij}$ due to the companion is then given to the leading post-Newtonian order by~\cite{Flanagan:2006sb}
\begin{equation}
\label{eq:Bijbinary}
    B_{ij} = \frac{6 M_2}{r^5} z_{(i}\epsilon_{j)kl}z_k\dot{z}_l
\end{equation}
where $r$ is the relative separation and we use lower-case Latin indices for the spatial components of tensors in this frame.

We further specialize to quasi-circular orbits of constant radius $\dot r=\ddot r=0$ and parameterize the orbit using two angles: the azimuthal orbital phase $\phi$ and the inclination angle $\psi$ of the spin axis of the NS relative to the orbital angular momentum such that the position vector becomes 
\begin{equation}
\label{eq:orbitalcoords}
    {\bm z}(t) = r(\cos\psi  \cos\phi(t),\, \sin\phi(t),\, \sin\psi  \cos\phi(t)). 
\end{equation}
The spin inclination angle $\psi$ is often approximated as constant because its change is very small~\cite{Kidder:1995zr,Ma:2020oni}. 
The transformation of~\eqref{eq:Bijbinary} from the NS's center of mass frame to the co-rotating frame is given by
\begin{equation}
\label{eq:frametrans}
B_{IJ} = R_I^iR_J^jB_{ij},
\end{equation}
where $R_I^i$ are rotation matrices. We assume that the NS's spin is along the $z$-axis in the co-rotating frame such that ${\bm \Omega}=(0,0,\Omega)$ and  $R_{1}^1=R_{2}^2=\cos(\Omega t)$,$R_{1}^2=\sin(\Omega t)=-R_{2}^1$, $R_{3}^3=1$ with all other components vanishing. The body label $1$ on $\Omega$ is implied here. To reduce~\eqref{eq:effectivelovenumber} to a function of the orbital parameters also requires the steady-state solution of the oscillator equations of motion~\eqref{eq:relEOM}. This is most conveniently calculated in a spherical-harmonic basis, using that
\begin{equation}
\label{eq:decomposeQ}
Q_{\cal B}^{IJ} = N_2 \sum_m \mathcal{Y}_{IJ}^{2m} Q^{\cal B}_m,
\end{equation}
where $N_2 = \sqrt{8 \pi / 15}$ and $\mathcal{Y}_{IJ}^{2m}$ are symmetric-trace-free tensors whose components are complex numbers~\cite{1980RvMP...52..299T}. We use a similar decomposition as~\eqref{eq:decomposeQ} for $B_{IJ}$. The equations of motion~\eqref{eq:relEOM} can then be expressed as 
\begin{equation}\label{eq:relEOMlm}
 \ddot Q^{\cal B}_m + i m \Omega \hat\omega^{\cal B}\dot Q^{\cal B}_m = -\frac{8}{3}(\sigma_\text{irrot} - \sigma_\text{stat}) \dot B_m ,
\end{equation}
In order to solve~\eqref{eq:relEOMlm} for the case of interest here, we extract from~\eqref{eq:Bijbinary},~\eqref {eq:frametrans} and~\eqref{eq:orbitalcoords} the spherical harmonic components 
\begin{equation}
B_m = N_2 \mathcal{Y}^{*2 m}_{IJ} {B}_{IJ}, 
\end{equation} 
where the asterisk denotes complex conjugation.
For circular orbits, these coefficients are given by 
\begin{subequations}\begin{eqnarray}
\label{eq:Bmcomps}
2 B_2 e^{-2i\Omega t}&=&\bar {\cal B} (2i \sin\psi\sin\phi-\sin 2 \psi \cos\phi)\\
B_1e^{-i\Omega t}&=&\bar {\cal B}(i \cos\psi\sin\phi-\cos 2\psi\cos\phi)\qquad \\
B_0&=&\bar {\cal B}\sqrt{3/2}\,\cos\phi\sin 2\psi,
\end{eqnarray}
with 
\begin{equation}
    \bar{\cal B}=\frac{3 M_2 \omega}{r^2}, \quad \omega=\dot \phi .
\end{equation}
\end{subequations}
The results for negative $m$ are obtained from the relation
\begin{equation}
B_{-m}=(-1)^m B^*_m.
\end{equation} 
 
Using these forcing terms in the equations of motion~\eqref{eq:relEOMlm} and solving for steady state solutions for $Q^{\cal B}_m$  leads to 
\begin{subequations}
\label{eq:Qmsol}
\begin{eqnarray}
    Q_2e^{-2i\Omega t}&=&\frac{8 {\bar{\cal B}}\, (\sigma_\text{irrot} - \sigma_\text{stat})}{3\, D_{2}}\left[i{\cal A}_{2,s}c_\phi+C_{2,s}s_\phi\right], \quad\; \quad\\ Q_1e^{-i\Omega t}&=& \frac{8 {\bar{\cal B}}\, (\sigma_\text{irrot} - \sigma_\text{stat})}{3\, D_{1}}\left[i{\cal A}_{1,c}c_\phi+C_{1,c}s_\phi\right]\\
    Q_0&=& -4\sqrt{6}\frac{M_2(\sigma_\text{irrot} - \sigma_\text{stat})}{r^2}\sin 2\psi \sin\phi, 
    \end{eqnarray}
     where $c_\phi=\cos(\phi)$, $s_\phi=\sin(\phi)$ and
    \begin{eqnarray}
        {\cal A}_{m,s}&=& \omega \sin\psi + (1+\hat\omega^{\cal B})\Omega \sin 2\psi\\
        C_{m,s}&=& m\Omega (1+\hat\omega^{\cal B})\sin\psi+\frac{\omega}{m} \sin 2\psi,
    \end{eqnarray}
   with the corresponding quantities with subscripts $c$ obtained by replacing '$\sin$' by '$\cos$' in the above expressions. The denominators in~\eqref{eq:Qmsol} are given by 
    \begin{equation}
    \label{eq:Dmdef}
        D_m=\left[\omega - m\Omega (1+\hat\omega^{\cal B})\right]\left[\omega + m\Omega (1+\hat\omega^{\cal B})\right] \vspace*{0.5cm}.
    \end{equation}
\end{subequations}

The final step is to use these results to obtain the effective Love number. The relevant tensor contractions entering~\eqref{eq:effectivelovenumber} are given by 
\begin{equation}
\label{eq:contr}
{B}_{IJ}{B}^{IJ}=2\bar{\cal B}^2, \quad \dot Q_{IJ} B^{IJ}=\sum_{m=-2}^2 \dot Q_m  B_{-m}.
\end{equation}
Using~\eqref{eq:contr}  in~\eqref{eq:effectivelovenumber} leads to the instantaneous effective Love number. Performing an orbit-average for the case considered here amounts to
\begin{equation}
\label{eq:sigmaintegr}
  \sigma_\text{eff} 
    =  \sigma_{\rm stat}-\frac{3}{16\bar{\cal B}^2} \frac{\omega}{2\pi}\int_0^{2\pi/\omega} \sum_{m=-2}^2 \dot Q_m  B_{-m} dt.
\end{equation}
Substituting~\eqref{eq:Qmsol} and~\eqref{eq:Bmcomps} into~\eqref{eq:sigmaintegr} leads to the final result for the effective Love number for one of the bodies
 \begin{widetext}
 \begin{eqnarray}
\label{eq:sigmaeffbinary}
    \sigma_\text{eff} 
    &=&\sigma_{\rm stat} +\frac{3(\sigma_\text{irrot} - \sigma_\text{stat})}{8}(\sin 2\psi)^2+\frac{(\sigma_\text{irrot} - \sigma_\text{stat})}{2D_1 }\bigg\{\omega \Omega \hat\omega^{\cal B}\left(\cos\psi+\cos 3\psi\right)+\left[\omega^2-\Omega^2(1+\hat\omega^{\cal B})\right]\left(1+\cos\psi\cos 3\psi\right)\bigg\}\nonumber\\
    &&+\frac{(\sigma_\text{irrot} - \sigma_\text{stat})(\sin\psi)^2}{4D_2 }\left[8\omega \Omega \hat\omega^{\cal B}\cos \psi+\left(\omega^2-4\Omega^2(1+\hat\omega^{\cal B})\right)(3+\cos 2\psi)\right].
  \end{eqnarray}
\end{widetext}
In a binary system of two NSs, one must add the same contribution but with the parameters of the companion body.

\subsection{Features of the effective response}
\subsubsection{Effects of the spin orientation}
The poles of the response~\eqref{eq:sigmaeffbinary}, i.e. where one of the factors in the denominators given in~\eqref{eq:Dmdef} vanishes, correspond to the four different mode resonances for the $m\neq 0$ modes. 

For special cases of the spin inclination angle only a subset of the modes contributes to $ \sigma_\text{eff}$, as also evident from~\eqref{eq:Qmsol}. For example, for aligned spin corresponding to $\psi=0$ the response~\eqref{eq:sigmaeffbinary} reduces to
\begin{equation}
    \sigma_\text{eff}\mid_{\psi=0}=\sigma_{\rm stat}+\frac{(\sigma_\text{irrot} - \sigma_\text{stat})\left(\omega- \Omega\right)}{\omega-\Omega(1+\hat\omega^{\cal B})}
\end{equation}
This shows that for aligned spins, and within our approximations, the only pole in the response is $\omega \to  \Omega(1+\hat\omega^{\cal B}) $ which corresponds to the $m=1$ resonance frequency. 

Another special case is a spin inclination of $\psi=\pi/3$, where the contribution from the $|m|=1$ modes is non-resonant. This can be seen either from~\eqref{eq:Qmsol}, by noticing that the numerator in $Q_1$ for this special value of $\psi$ will involve factors of $\omega -\Omega (1+\hat\omega^{\cal B})$ which cancel the divergent term in the denominator, or by considering the third term in~\eqref{eq:sigmaeffbinary} showing the same effect.  

\subsubsection{Adiabatic limits}
\label{subsubsec:adiabaticresponse}
Above, we have computed the response assuming a fixed orbit, obtaining divergences in the response at the resonances. However, in a binary inspiral, the continued GW dissipation causes the system to evolve through the resonance, exciting the mode amplitudes only to a finite maximum value. This effect was already examined in detail in~\cite{Flanagan:2006sb}, who also developed an effective waveform model for these resonance-induced effects. A missing phenomenon from these and subsequent studies were the additional adiabatic effects due to the behavior of the modes far from the resonances. To compute the relevant NS parameters characterizing the adiabatic response, we consider the asymptotic limits of $\sigma_\text{eff}$ long before or after a resonance. The subtleties with extracting the relevant limits were discussed in detail in~\cite{Gupta:2020lnv}, as the appropriate ordering of limits between $\omega, \Omega \to 0$ is delicate and depends on the situation. In particular, the relevant adiabatic limit before the mode resonance is obtained by considering $\omega\to 0$ in~\eqref{eq:sigmaeffbinary}, while the post-resonance adiabatic limit is given by taking the limit $\Omega \to 0$ first. This leads to the asymptotic expressions pre- and post-resonance respectively
\begin{equation}
\label{eq:adiabaticlimits}
\sigma^{\rm asym}=\begin{cases}\sigma_{\rm stat}+\frac{(\sigma_{\rm irrot}-\sigma_{\rm stat})[8+3\hat\omega_{\cal B}\sin (2\psi)^2]}{8(1+\hat\omega_{\cal B})} \\
\sigma_{\rm irrot} \qquad \qquad
\end{cases} .
\end{equation}
We will use the above insights into the features of the response to assemble an approximate waveform model that properly accounts for both resonance and adiabatic effects. 

\section{Effective waveform model with adiabatic and resonance effects}
\label{sec:waveform}

\subsubsection{Approximate waveform model}
Computing the impact of gravitomagnetic tidal effects on the GW signals from inspiraling NS binary systems is a complicated task. Here, we bypass these challenges by assmebling a simple effective model for the gravitomagnetic imprints in frequency-domain descriptions of the GW signals based on adapting existing results using the insights developed in the previous section. Such a model is very useful for scoping out the features, magnitude, and consequences of the various gravitomagnetic effects in future GW measurements, and for identifying focus areas for more detailed modeling.  In addition to the gravitomagnetic effects, we also include the dominant adiabatic gravitoelectric tidal effects to understand the impacts on the overall information on NS matter.

Specifically, we write the GW phasing in the frequency domain as
\begin{subequations}
\begin{equation}
    \label{eq:phasecontibutionfromtidal}
    \Psi=2\pi f t_c-\phi_c+\Psi_{\rm pm}+\Psi^{\rm tidal}_{\rm ad}+\Psi^{\rm tidal}_{\rm res},
\end{equation}
where $t_c$, and $\phi_c$ are the reference time and phase and $f$ is the GW frequency. The term $\Psi_{\rm pm}$ is the point-mass contribution, for which we use the post-Newtonian TaylorF2 results given e.g. in Eq.~(3.18) of~\cite{Buonanno:2009zt}. For the adiabatic tidal contributions $\Psi^{\rm tidal}_{\rm ad}$ we use the results of~\cite{Hinderer:2007mb, Flanagan:2007ix,Vines:2011ud, Damour:2012yf,Henry:2020ski,Yagi:2013sva,JimenezForteza:2018rwr,Banihashemi:2018xfb} given by 
\begin{eqnarray}
\label{eq:tidalphasing}
  && \Psi^{\rm tidal}_{\rm ad}= - b_0\tilde \Lambda f^{5/3} + \left(-b_1 \tilde \Lambda+b_2 \delta \tilde \Lambda+ b_3\tilde \Sigma\right) f^{7/3}\qquad\nonumber\\
   &&\qquad- b_5\hat \Sigma(\chi_1,\chi_2) f^{8/3}+f_2 f^{8/3}+f_3 f^3+f_4 f^{10/3}, 
\end{eqnarray}
and take the resonance-induced effects from~\cite{Flanagan:2006sb, Ma:2020oni} in the form
\begin{equation}
\label{eq:phaseshift}
     \Psi^{\rm tidal}_{\rm res}=\quad - \sum_{i=1,2}\left( 1-\frac{f}{f^{\rm res}_{i}}  \right) |\Delta \Phi_i| \Theta(f-f^{\rm res}_{i}).
\end{equation}
\end{subequations}
We note that the signs of all the contributions made explicit here correspond to those relevant for the parameter choices for the case studies discussed in Sec.~\ref{sec:results} below, with all the tidal parameters $\tilde \Lambda, \delta \Lambda, \tilde \Sigma, \hat\Sigma$ defined below being positive. We also see that the resonance contribution is a  distinct sudden
change in the phase and time of the GW signal at the resonance, whose scaling with the frequency is degenerate with that of the gauge parameters $\phi_c$ and $t_c$ in the phasing~\eqref{eq:phasecontibutionfromtidal}. The various coefficients in~\eqref{eq:tidalphasing} are given by 
\begin{eqnarray}
b_0&=&\frac{117 (\pi M)^{5/3}}{256\nu}, \quad b_1=\frac{9345 (\pi M)^{7/3}}{8192\nu}\\
b_2&=&\frac{19785 (\pi M)^{7/3}}{46592\nu}\sqrt{1-4\nu}\\
b_3&=&\frac{3 (\pi M)^{7/3}}{128\nu}=b_4/(\pi M)^{1/3},
\end{eqnarray}
with $M=M_1+M_2$ the total mass and $\nu=M_1 M_2/M^2$.
The functions $f_j$ depend on $\tilde \Lambda, \delta \tilde \Lambda$, and for $f_2$ additionally on the spins $\chi_{1,2}$. In particular, the expression for the function $f_2$ in~\eqref{eq:tidalphasing} is obtained from Eqs.~(7) and (9) of~\cite{Jimenez-Forteza:2018buh} and those for $f_3, f_4$ from Eq. (6.6b) of~\cite{Henry:2020ski}.  
 The parameters $\tilde \Lambda$ and $\delta \Lambda$ characterizing the gravitoelectric effects are given by 
\begin{eqnarray}
 \tilde\Lambda &=&\frac{16}{13}\left(\frac{12}{X_1}-11\right) X_1^5\Lambda_1+
  (1\leftrightarrow 2)\,,\label{defLambda} \\
2 \delta\tilde\Lambda&=&\sqrt{1-4\nu} \left(1-\frac{13272}{1319}\nu+ \frac{8944}{1319}\nu^2\right)(\Lambda_1+\Lambda_2)\\
  &+&\left(1-\frac{15910}{1319}\nu+\frac{32850}{1319}\nu^2+\frac{3380}{1319}\nu^3\right)(\Lambda_1-\Lambda_2)\nonumber
  \end{eqnarray}
 with $
\Lambda_{i}$ the dimensionless quadrupolar gravitoelectric tidal deformability parameters of each body indexed here by $i$. We also denote $X_{i} = M_{i}/M$ and $(1\leftrightarrow 2)$ indicates the operation of adding the same terms but with the body labels interchanged. 
The gravitomagnetic parameters in~\eqref{eq:tidalphasing} are defined by~\cite{Jimenez-Forteza:2018buh} 
\begin{subequations}
\label{eq:gravitoSigma}
\begin{eqnarray}
    \tilde \Sigma &=& \left(\frac{6920}{7}-\frac{20740}{21X_1}\right)X_1^5\Sigma_1 +(1\leftrightarrow 2)
\\
    \hat{\Sigma} &=& \left[\chi_1-\left(\frac{4933}{3X_1}-\frac{9865}{3}+ 1644X_1\right)\chi_2\right]X_1^5\Sigma_1\nonumber\\
    &&+(1\leftrightarrow 2),
\end{eqnarray}
with $\chi_{i} = S_{i}/M_{i}^2$ the dimensionless spin parameter of each body. We use for the dimensionless gravitomagnetic deformability parameters $\Sigma_i$ the asymptotic results of Sec.~\ref{subsubsec:adiabaticresponse} to replace
\begin{equation}
\Sigma_{i}=\frac{\sigma^{\rm asym}_{i}}{M_{i}^5}
\end{equation}
\end{subequations}
using the appropriate pre- or post-resonance expressions from~\eqref{eq:adiabaticlimits}.

In the resonance contributions~\eqref{eq:phaseshift}, the quantity $\Theta$ denotes the Heaviside step function, $f^{\rm res}_{i}$ are the GW frequencies at which the mode resonances occur. They are related to the gravitomagnetic mode frequencies $\omega_{2m}^\mathcal{B}$ by
\begin{equation}
f^{\rm res} =\frac{\omega_{2m}^{\rm inertial}}{\pi} ,
\label{eq:f0gen}
\end{equation}
where the mode frequencies in the inertial frame can be obtained by shifting
\begin{equation}
\omega_{2m}^{\rm inertial}= \omega_{2m}^\mathcal{B} - m\Omega. 
\end{equation}
The quantities $\Delta \Phi_{1,2} $ are the corresponding resonance-induced phase shifts, which, for the $l=2$ modes with $m=2$ and $m=1$, are given by~\cite{Flanagan:2006sb} 
\begin{subequations}
\label{eq:phasejump}
\begin{equation}
 \Delta \Phi_{2m} = -\frac{10\pi^2}{192} \left(\frac{2m}{3}\right)^{2/3}(M_i\Omega_i)^{2/3} \left(\frac{M_i}{\mathcal M}\right)^{10/3}\mathcal I_i^{2m},  
\end{equation}
where ${\cal M}=(M_1 M_2)^{3/5}/M^{1/5}$ is the chirp mass and 
\begin{eqnarray}
 \mathcal I_i^{22} &=& (\bar{I}_i^r)^2  \sin^2(\psi_i)\cos^4\left(\frac{\psi_i}{2}\right)\left( 1 - X_i\right)\\
 \\
 \mathcal I_i^{21}&=& (\bar{I}_i^r)^2  \cos^2\left(\frac{3\psi_i}{2}\right)\cos^2\left(\frac{\psi_i}{2}\right)\left( 1 - X_i\right),\qquad
\end{eqnarray}
with $\bar{I}_i^r$ related to the dimensionless relativistic tidal deformabilities by~\cite{Gupta:2020lnv}
\begin{equation}
\label{eq:IcalB}
 (\bar{I}^r)^2 = \frac{15}{4 \pi} ( \Sigma_\text{stat} - \Sigma_\text{irrot} ).
\end{equation}
\end{subequations}

\subsubsection{Reducing the number of matter parameters using quasi-universal relations}
Even within the restricted context considered here, the effective GW model for the tidal signatures~\eqref{eq:tidalphasing} and~\eqref{eq:phaseshift} contains ten matter parameters, namely the deformabilities $\Lambda_i, \sigma_i^{\rm stat}, \sigma_i^{\rm irrot}$ and resonance frequencies for the $m=1$ and $m=2$ modes for each body. Such a large number of extra parameters prevents the data analysis from yielding meaningful results. We reduce the number of parameters by using empirical quasi-universal relations that are approximately independent of the equation of state and enable an approximate reduction of the matter parameters to one deformability $\Lambda$ for each body.  
The quasi-universal relations are of the form~\cite{Yagi:2013sva,JimenezForteza:2018rwr} 
\begin{equation} 
\ln(\mp\Sigma ) = \sum_{n=0}^5 a_n Y^n, 
\end{equation}
with the irrotational case corresponding to the minus sign and coefficients $a_n^{\rm irrot}=\{-2.03, 0.487, 0.00969 , 0.00103 , 9.37 \times 10^{-5} , 2.24 \times 10^{-6}\}$, while the plus sign applies for the static case with 
 $a_n^{\rm stat}=\{  -2.66 , 0.786,-0.01, 0.00128 ,- 6.37 \times 10^{-5}, 1.18 \times 10^{-6} \}$ and
where 
\begin{equation}
    Y=\ln(\Lambda).
\end{equation}

The GW frequencies appearing in the resonant mode contributions~(\ref{eq:phaseshift}) are given by~\eqref{eq:f0gen}, which can be written explicitly as
\begin{equation}
f^{\rm res} = \frac{1}{\pi}(\kappa_m-m)\,\Omega, 
\label{eq:f0}
\end{equation}
where the parameter $\kappa_m$ reduces to $\kappa_m\to 2m/3$ in the Newtonian limit, while for relativistic stars, it is approximately related to $\Lambda$ by~\cite{Idrisy:2014qca,Gupta:2022qgg}
\begin{equation}
  \label{eq:kappafit}
  \kappa_2 = 0.3668 + 0.0498 Y- 0.0025 Y^2,
\end{equation}
We note that these results from~\cite{Idrisy:2014qca} are specialized to the $m=2$ mode. Within the effective action model~\eqref{Lmag} we are using, the modes with different $m$ all have the same scaled frequency $\hat \omega_{\cal B}$ and hence the same $\kappa$. Thus, we use~\eqref{eq:kappafit} also for the $m=1$ modes. 

With the GW phasing model for the gravitomagnetic effects in hand, we next apply it in a data analysis framework to study the impact on GW measurements.
\section{Analysis framework}
\label{sec:gaussian}

A Bayesian data analysis framework is commonly used for GW signals, as explained e.g. in~\cite{LIGOScientific:2019hgc} and briefly reviewed below. We assume that in the absence of any GW signal, the detector noise $n$ has a Gaussian distribution, where louder noise realizations are less likely. In the presence of a signal $h$ with parameters ${\bm \theta}$, the data $d$ from the detector output can be decomposed as
\begin{equation}
d = h({\bm \theta})+n,
\end{equation}
for some noise realization $n$.  
Then, the  likelihood ${\cal L}$ for the detector to measure the data $d$ for a signal with parameters ${\bm \theta}$ is given by
\begin{equation}
    \text{log}\mathcal{L}(d|{\bm \theta}) = -\frac{1}{2}(d-h({\bm \theta})|d-h({\bm \theta})). \label{eq:LLfirst}
\end{equation}
Here, the meaning of $(.|.)$ differs on both sides of the equation: on the left hand side, $\mathcal{L}(d|{\bm \theta})$ denotes the conditional probability of observing the data $d$ for a collection of signal parameters ${\bm \theta}$, while on the right hand side, the notation $(.|.)$ indicates an inner product on the vector space of signals. For two signals $h_1$ and $h_2$ this inner product is defined as
    \begin{equation}
        (h_1|h_2) = 4 \mathcal{R} \int_{f_{\rm low}}^{f_{\rm high}} \frac{\tilde{h}_1(f)^*\tilde{h}_2(f)}{S_n(f)}df .
    \end{equation}
    The symbol $\mathcal{R}$ denotes the operation of taking the real part, the integration limits are the lower and upper frequency range considered, $S_n$ is the noise spectral density of the detector, and the tilde and asterisk indicate the Fourier transform and complex conjugate respectively. 
This log likelihood~\eqref{eq:LLfirst} can be further expanded as
\begin{equation}
    \label{eq:likelihoodfunction}
    \text{log}\mathcal{L}(d|{\bm \theta}) = -\frac{1}{2}\left[(d|d)+(h({\bm \theta})|h({\bm \theta}))-2(d|h({\bm \theta}))\right].
\end{equation}
The first term is proportional to the log noise evidence and the second term $(h({\bm \theta})|h({\bm \theta}))=\rho^2_{\rm opt}$ is called the optimal matched filter signal-to-noise ratio (SNR) squared. The third term is the product of the optimal SNR and the matched filter SNR given by $(d|h({\bm \theta}))=\rho_{\rm opt}\rho_{\rm mf}$.

 The posterior probability distribution of the parameters $\bm \theta$ follows from Bayes’ theorem:
\begin{equation}
    \label{eq:bayestheorem}
    p({\bm \theta} | \mathcal{H}, d, \mathcal{I}) = \frac{p(d|\mathcal{H}, {\bm \theta}, \mathcal{I}) p({\bm \theta} | \mathcal{H}, \mathcal{I})}{p(d| \mathcal{H}, \mathcal{I})}
\end{equation}
where $\mathcal{I}$ is the background information, $\mathcal{H}$ is the hypothesis, i.e. the waveform model. The quantity $p({\bm \theta} | \mathcal{H}, \mathcal{I})$ is the prior probability, i.e. knowledge about the parameters within the model before analyzing the data,  $p(d| \mathcal{H}, \mathcal{I})$ is the evidence and $p(d|\mathcal{H}, {\bm \theta} , \mathcal{I})$ is the likelihood function which is identified with~\eqref{eq:likelihoodfunction}. 
Computing the posterior probability distribution of the parameters ${\bm \theta}$ requires  Markov chain Monte Carlo (MCMC) samplers~\cite{Foreman-Mackey:2012any}. 

The above framework is general but also computationally intensive, especially when taking into account the following considerations. Gravitomagnetic tidal effects are subdominant, though expected to be relevant for next-generation GW detectors. The detectors will have a much wider frequency band than current detectors such that signals from NS binaries will linger for many hours to days within the sensitive band. The associated tremendous computational costs severely limit the scope of explorative studies possible with the current MCMC code infrastructures. However, 'golden' events  similar to GW170817, which would have an SNR of over a thousand in next-generation detectors, will provide rich science yields, especially when combined with the larger number of  events with lower SNR. For the exploratory studies in this paper, we use an MCMC analysis in a lower-dimensional subspace of the signal parameters, which we validate against a simplified data analysis framework based on approximations for large SNR: the Fisher Matrix formalism. 
For a high SNR event and Gaussian noise, the probability distributions of the best-fit parameters will be Gaussians centered around the actual values.
Let ${\bm \theta}$ be the true value of the parameters and ${\bm \theta} + {\bm \Delta \theta}$ the best-fit parameters in the presence of  Gaussian noise. Then for large SNR, the likelihood function is given by
\begin{equation}
p({\bm \Delta \theta}) = \mathcal{N}e^{-\frac{1}{2}\Gamma_{ij}\Delta \theta^i \Delta \theta^j}
\end{equation}
where the Fisher matrix $\Gamma_{ij}$ is defined as,
\begin{equation}
      \Gamma_{ij} = \bigg( \frac{\partial h}{\partial \theta^i} \bigg| \frac{\partial h}{\partial \theta^j} \bigg).
\end{equation}
 The 1-sigma error $\sigma^i$ on the parameters $\theta^i$ is then given by 
\begin{equation}
     \sigma^i = \sqrt{(\Gamma^{-1})^{ii}} .
\end{equation}

\section{Results}
\label{sec:results}
We use the analysis frameworks described in Sec.~\ref{sec:gaussian} to analyze the impact of gravitomagnetic tides on the measurability of the tidal Love number $\Lambda$. For simplicity, we focus on the Cosmic Explorer (CE) detector~\cite{LIGOScientific:2016wof} , however, we expect similar results for the Einstein Telescope~\cite{Punturo:2010zza}.

\subsection{Setup and parameter choices for case studies}
\label{subsec:parameters}
We consider a few illustrative cases for our analysis. These examples represent only a small subset of the expected range of diverse events but nevertheless yield useful insights. Specifically, we consider binary neutron stars with masses 
$
    (M_1, M_2) = (1.5, 1.3)\,M_\odot
$ and explore two values of the dimensionless spin parameters $\chi = 0.005$ and $\chi= 0.01$ for each NS, where $\chi$ refers to the spin magnitudes. For the tidal deformability  parameters we choose $(\tilde \Lambda, \delta \tilde\Lambda ) = (519, 48)$~\cite{LIGOScientific:2018cki}, corresponding to the MPA1 equation of state. We use quasi-universal relations~\cite{Yagi:2013bca} between the moment of inertia and $\Lambda$ to convert from $\chi$ to the spin frequency $\Omega$. In general, both the $m=1$ and $m=2$ resonances will contribute to the signals. To isolate each of these resonance effects and analyze its contributions, we choose spin inclination angles of $\psi=0$ (aligned spins) and $\psi=\pi/3$ such that only the $m=1$ or $m=2$ modes respectively undergo a resonant excitation within our approximations. We assume the same spin magnitudes and orientations for both NSs.

We analyze the signals in the CE detector sensitivity~\cite{LIGOScientific:2016wof} between $f_{\rm low} = 5$Hz and $f_{\rm high}\sim 1720$Hz, which is a proxy for the merger frequency based on the estimates for nonspinning NSs from~\cite{Dietrich:2017aum}. Unless otherwise specified, the SNR for the signals from these systems is 1800 for the CE detector, which corresponds to an event similar to GW170817. 

\begin{figure}[H]
\centering
    \includegraphics[scale=0.405]{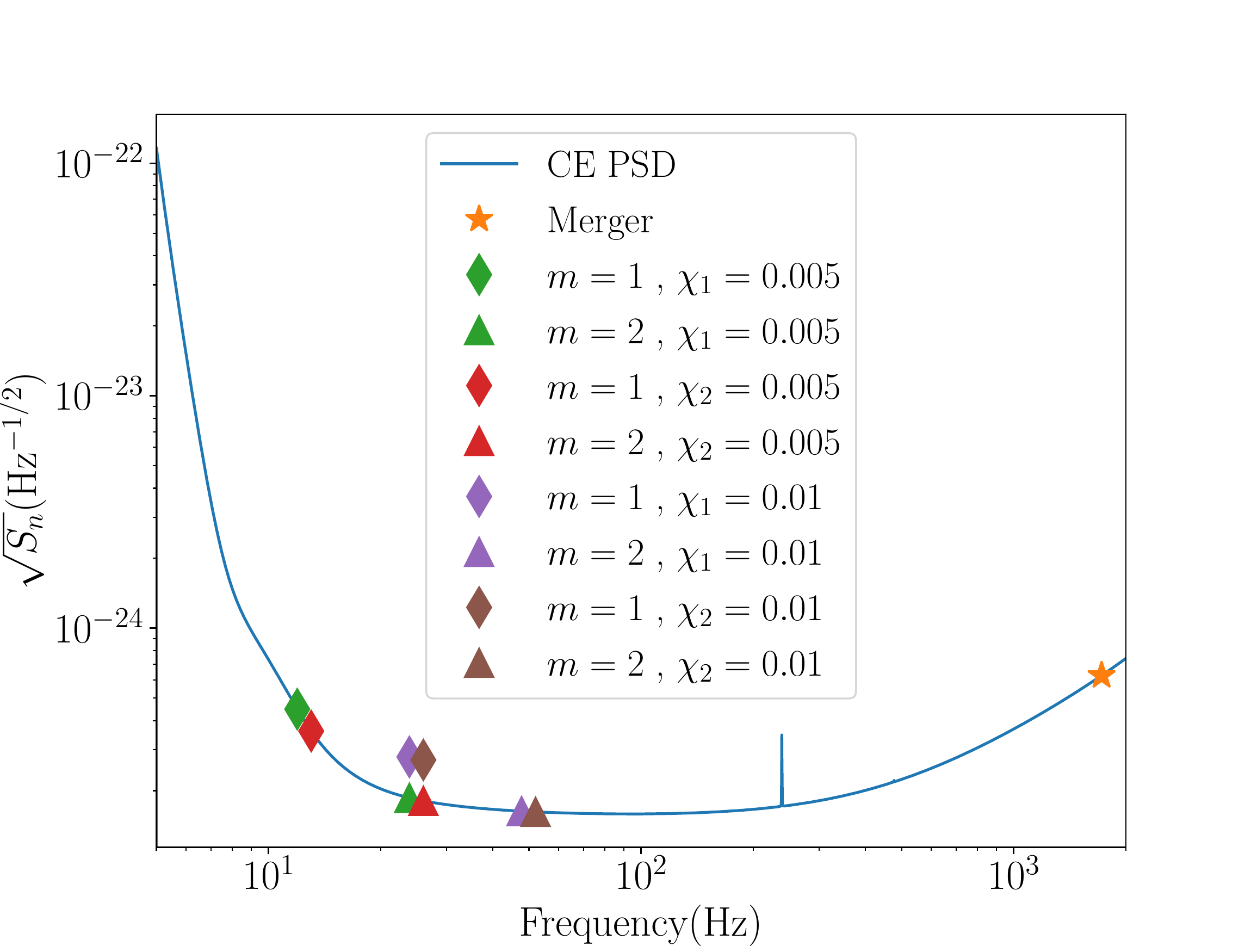}
  
    \caption{\emph{CE noise spectral density and various mode resonances} for the two bodies and varying spins $\chi$. Green and purple symbols refer to the NS with mass $M_1=1.5M_\odot$ with lower and higher spin respectively, while red and brown symbols are the corresponding values for the companion of mass $M_2=1.3M_\odot$. Diamond shapes denote the modes with azimuthal number $m=1$, triangles those with $m=2$.}
    \label{fig:PSD}
\end{figure}

For the above choices of binary parameters, the mode resonance frequencies for the larger and smaller mass NSs are given by $f_1^{\rm res}=$12 (24)Hz and $f_2^{\rm res}=$13 (26)Hz for the $m=1$ ($m=2$) modes respectively and taking the spin magnitudes to be $\chi=0.005$; they increase to twice these numbers when doubling the spin magnitudes to $\chi=0.01$. Figure~\ref{fig:PSD} illustrates the location of these resonance together with the power spectral density of the CE detector~\cite{LIGOScientific:2016wof}.

To study the consequences of different effects, we consider different tidal waveform models. We refer to the 'PNTidal' model as the piece of~\eqref{eq:tidalphasing} involving only the adiabatic gravitoelectric tidal effects characterized by $\tilde\Lambda, \delta \Lambda$, and denote models that also include gravitomagnetic effects by  PNTidal$^\text{modes}$ for the resonant contributions~\eqref{eq:phaseshift}, PNTidal$_\text{asym}$ for the asymptotic adiabatic contributions, and PNTidal$_\text{asym}^\text{modes}$ for the model which includes all gravitomagnetic effects. Because we work only to linear order in the spins, we neglect the effects of spin-induced multipole moments on the GWs. 

\subsection{Consistency checks}
\label{subsec:checks}
\subsubsection{Fisher matrix versus Bayesian parameter estimation and effect of the dimensionality of the parameter space}

\begin{figure}[H]
    \centering
    \includegraphics[scale=0.41]{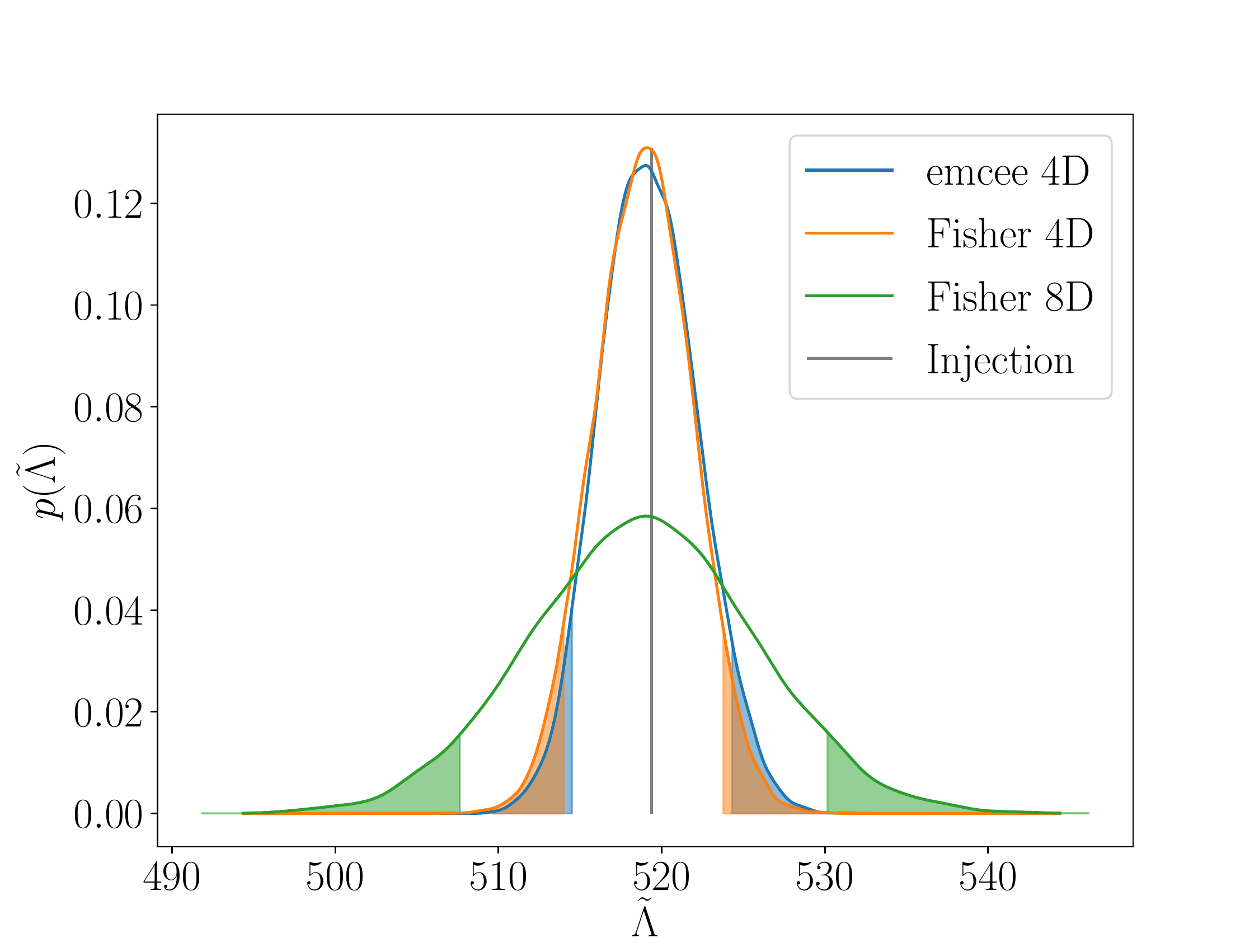}
    \caption{\emph{Posterior probability distribution of $\tilde \Lambda$ for SNR 1800 with the PNTidal waveform model (without gravitomagnetic effects) used for injection and recovery}. The label 4D refers to a reduced parameter space of the tidal deformabilities $\tilde \Lambda, \delta\tilde\Lambda$ and the time and phase of coalescence $t_c,\phi_c$ with all other parameters fixed, while 8D also includes the sampling of the mass and spin parameters for each body. The results from the Fisher matrix (orange curve) agree well with the corresponding Bayesian analysis (blue curve), with both centered on the injected value (vertical line). The green curve shows the broadening of the distribution when doubling the dimensionality of the parameter space sampled.  
The shaded tails of the curves indicate regions outside the 90$\%$ credible interval. For the parameter $\delta \tilde \Lambda$, both the  4D and 8D posteriors are essentially flat in this case. }
    \label{fig:fisher}
\end{figure}

The Fisher matrix approximation is valid for high SNR, which we expect to hold for most of the case studies considered here. To assess the validity of this expectation we compare with Bayesian parameter estimation results for the case of the PNTidal matter model. In principle, the waveforms are characterized by 17 parameters, after reducing the matter parameters to just $\Lambda$ for each body. Exploring the full parameter space is thus very computationally expensive. For efficiency, we focus the comparitive analysis here only on the following restricted subset the intrinsic parameters  
\begin{equation}
    \label{eq:4freeparameter}
    \theta = (t_c, \phi_c, \tilde \Lambda, \delta \tilde\Lambda )
\end{equation}
and fix the other parameters to be $\psi_1=0$, $\psi_2=0$, $\chi_1 = 0.01$ $\chi_2=0.01$. This subset was chosen to contain the matter-related parameters $\tilde \Lambda$ and $\delta \tilde \Lambda$ as well as $t_c$ and $\phi_c$ which are degenerate with mode resonance effects. We sample the Fisher likelihood with the prior constraints $\Lambda_1 \geq 0 , \Lambda_2 \geq 0 $. We also perform a Bayesian analysis for the same setup using the \texttt{emcee} sampler to obtain the posterior probability distribution of the parameters. Figure~\ref{fig:fisher} shows the results of both analyses. 
We see that in this case, the results from the Fisher (blue curve) and Bayesian (orange curve) frameworks agree well and are centered on the injected value (vertical line). To obtain an estimate of the changes in the width of the posterior distributions when including more parameters, in particular the masses and spin magnitudes for each body, we also perform a Fisher analysis for eight free parameters ${\bm \theta} = (t_c, \phi_c, M_1, M_2, \tilde \Lambda, \delta \tilde\Lambda, \chi_1, \chi_2)$. More specifically, we obtain a mean and 90 percentile results of
$\tilde \Lambda=519^{+5.1}_{-4.7}$ from the MCMC and $\tilde \Lambda=518.9^{+4.8}_{-4.9}$ from the Fisher analyses with four free parameters respectively, which shows that they are in good agreement. For an eight-dimensional parameter space we find $\tilde \Lambda=518.9^{+11.3}_{-11.2}$, which indicates that when doubling the dimensionality of the parameter space the posterior distributions increase in width by about a factor of two.  
The good agreement between the Fisher and Bayesian results also provides a useful check of the 4D MCMC sampling, which is the method we will continue to use in what follows. 
\subsubsection{Comparison to the adiabatic effects studied in~\cite{Jimenez-Forteza:2018buh}}
\label{subsec:comparePani}
\begin{figure}[H]
    \centering
    \includegraphics[scale=0.4]{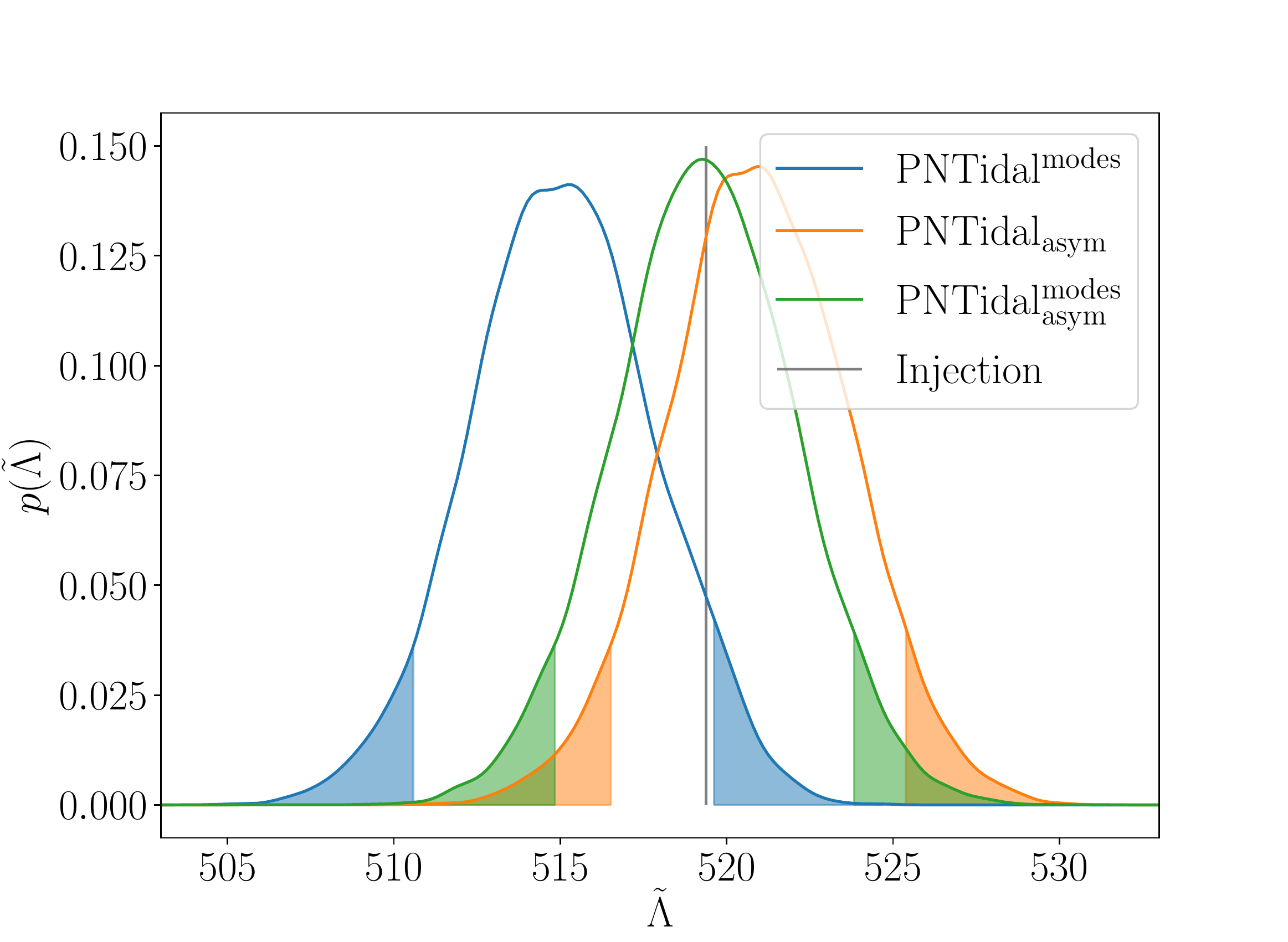}
    \caption{\emph{Shifts in the posterior distribution for $\tilde \Lambda$ due to adiabatic and resonant gravitomagnetic effects}. This case study is for SNR 1800, aligned spins $\chi=0.005$ and sampling only on $(\tilde \Lambda, t_c, \phi_c)$ with all other parameters fixed.  We inject with a waveform that includes all effects PNTidal$_\text{asym}^\text{modes}$ and recover with the same waveform (green curve) and those that include only the resonance jumps (blue curve) and only the adiabatic effects (orange curve). In this case the contribution from the adiabatic effects is dominant; omitting them (as for the results shown by the blue curve) leads to the largest shifts in the distribution.}
    \label{fig:comparePani}
\end{figure}

The final consistency check we perform here is to compare with the results of~\cite{Jimenez-Forteza:2018buh} for the impact of adiabatic gravitomagnetic effects on measurements of $\tilde\Lambda$. Following~\cite{Jimenez-Forteza:2018buh} we restrict our analysis to only three free parameters
$
    \theta = (t_c, \phi_c, \tilde \Lambda),
$
with all the other parameters fixed.  Figure~\ref{fig:comparePani} shows the results for aligned spins of magnitude $\chi=0.005$. Comparing the orange curve (no mode resonances) and blue curve (no adiabatic effects) to the green curve shows that in this case the largest impact of gravitomagnetic effects is due to the adiabatic limits, while mode resonances play a subdominant role.
Specifically, we obtain $\tilde\Lambda=519.3_{-4.5}^{4.5}$ with the full model (green curve) that was also used for the injection and thus quantifies the statistical errors. Using only the adiabatic effects (orange curve) leads to   $\tilde\Lambda=520.9^{4.4}_{-4.4}$, which is close to the injected value. On the other hand, when including only the mode resonances for the recovery while neglecting the adiabatic effects (blue curve) leads to a distribution that is  significantly shifted away from the injected value with $\tilde\Lambda=515.1^{4.6}_{-4.5}$.
The smallness of the effect of the mode resonances on measurements of $\tilde \Lambda$ is in part due to the fact that the resonance-induced phase corrections~\eqref{eq:phaseshift} have a scaling in frequency degenerate with the gauge parameters $t_c$ and $\phi_c$ in the phase~\eqref{eq:phasecontibutionfromtidal}, which absorbs some of the resonance effects into shifts of $t_c$ and $\phi_c$. The adiabatic effects show a similar magnitude as found in~\cite{Jimenez-Forteza:2018buh} based on only the irrotational or static Love numbers, which lead to shifts in the posterior distributions to lower and higher values respectively, c.f. Fig.~6 therein. While the specific choices for the case study here differ from~\cite{Jimenez-Forteza:2018buh} the setup is similar enough to interpret qualitative trends by comparing their findings to the adiabatic results represented by the orange curve in Fig.~\ref{fig:comparePani}, which uses the more realistic asymptotic Love numbers from~\eqref{eq:adiabaticlimits} as opposed to only $\sigma_{\rm irrot}$ for the entire waveform. 

\subsection{Physical effects}
Having performed the consistency checks discussed above, we next analyze the impact of various physical effects and parameter dependencies by sampling on the four-dimensional parameter space~\eqref{eq:4freeparameter}. We first consider nonspinning systems, where there is no effect from the mode resonances, then aligned spins with only the $m=1$ modes resonant, and finally spin orientations that maximize the effects of the $m=2$ modes. 

\subsubsection{Gravitomagnetic effects for nonspinning systems}
\label{subsect:nonspinning}
\begin{figure}[H]
    \centering
    \includegraphics[scale=0.65]{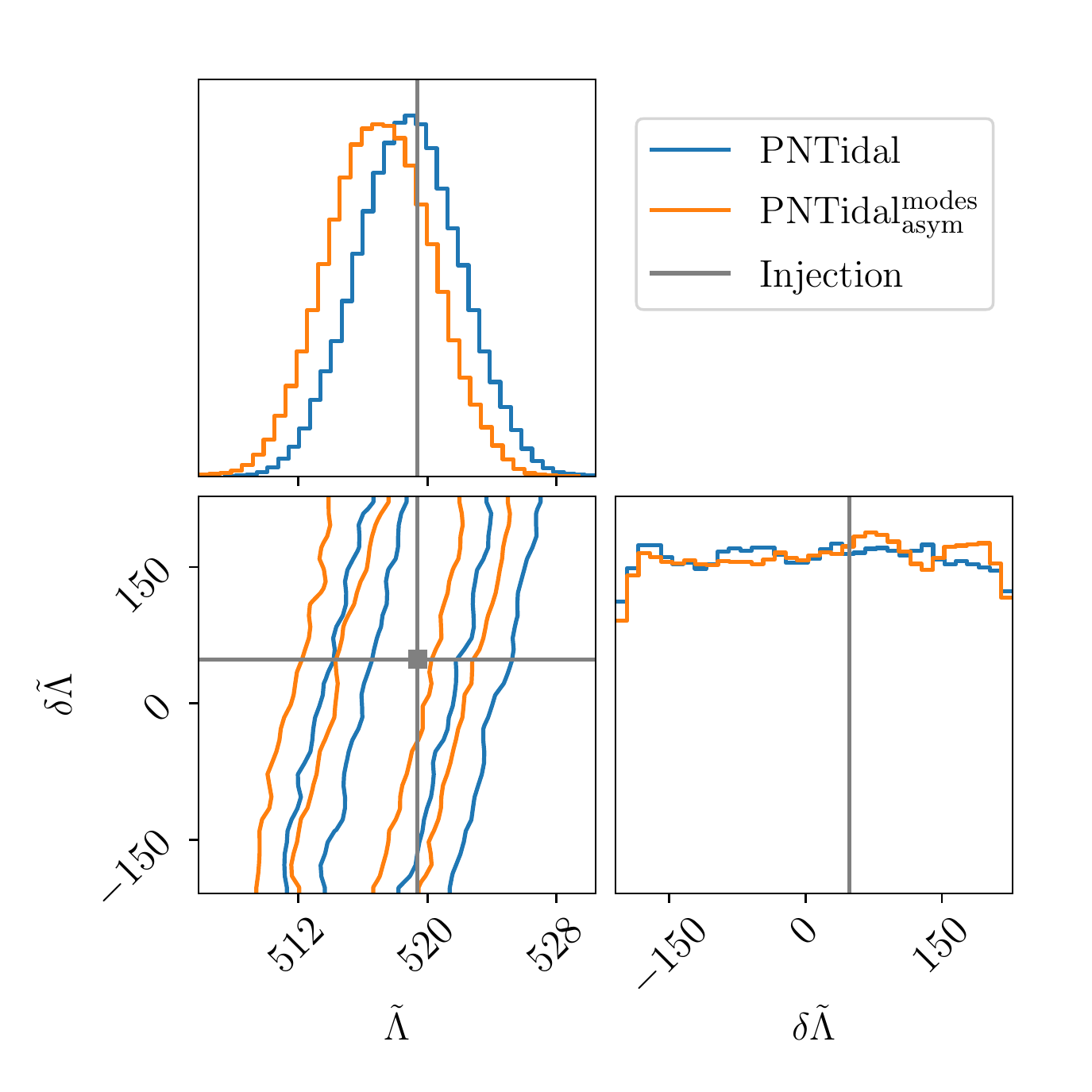}
    \caption{\emph{Posterior distributions of the tidal parameters for nonspinning NSs} at SNR 1800. The injection neglected gravitomagnetic tides, and the blue curve illustrates the recovery with the same waveform. The effect of gravitomagnetic tides, which are purely adiabatic in this case, is indicated by the orange curve. In the two-dimensional representation in the lower-left panel, the contours correspond to the one and two sigma confidence levels.}
    \label{fig:recover_Pv2_nospin}
\end{figure}

 For this study we use the PNTidal model without the gravitomagnetic effects as the reference baseline for the injection and set $\chi=0$. Figure~\ref{fig:recover_Pv2_nospin} shows the results for the posterior distributions in the tidal parameters $\tilde \Lambda$ and $\delta \tilde \Lambda$, with the two-dimensional representations given in the lower left panel, and the one-dimensional projections for each parameter in the upper and right panels. The one-dimensional representations are the full distributions, while the contours in the $\tilde\Lambda - \delta \tilde \Lambda$ plane correspond to the credible intervals at the one (68\%) and two (95\%) sigma confidence level. The blue curves in Fig.~\ref{fig:recover_Pv2_nospin} represent a consistency check that when injecting and recovering with the same model the mean is centered on the injected value indicated by the gray lines and quantify the statistical uncertainties. The orange curve in Fig.~\ref{fig:recover_Pv2_nospin} corresponds to the results obtained when including all gravitomagnetic effects, where, however, for nonspinning systems there is only an adiabatic gravitomagnetic effect, no resonances. As expected, we see that they induce a small shift in the posterior for $\tilde \Lambda$. We note that the difference to the study in Sec.~\ref{subsec:comparePani} is the model used for the injection, the value of the spins, which also impacts the adiabatic gravitomagnetic parameters~\eqref{eq:gravitoSigma}, and the dimensionality of the parameter space sampled. We also observe that the adiabatic effects have no significant impact on the measurability of $\delta \Lambda$ in this case, as the shape of the error ellipses and the flat distribution in $\delta\tilde \Lambda$ remain largely unaffected. 

\subsubsection{Effect of gravitomagnetic tides for aligned spins}
\label{subsec:alignedspinsall}
A more realistic scenario is to include finite spins of the NSs. We first consider the case of aligned spins, where the $m=1$ mode resonances contribute in addition to the adiabatic effects. As in Sec.~\ref{subsect:nonspinning}, we use the model without gravitomagnetic tides as the reference baseline for the injection and recover with the the same model (blue curve) as well as the model including all gravitomagnetic effects (orange curve). 
\begin{figure}[H]
    \includegraphics[scale=0.635]{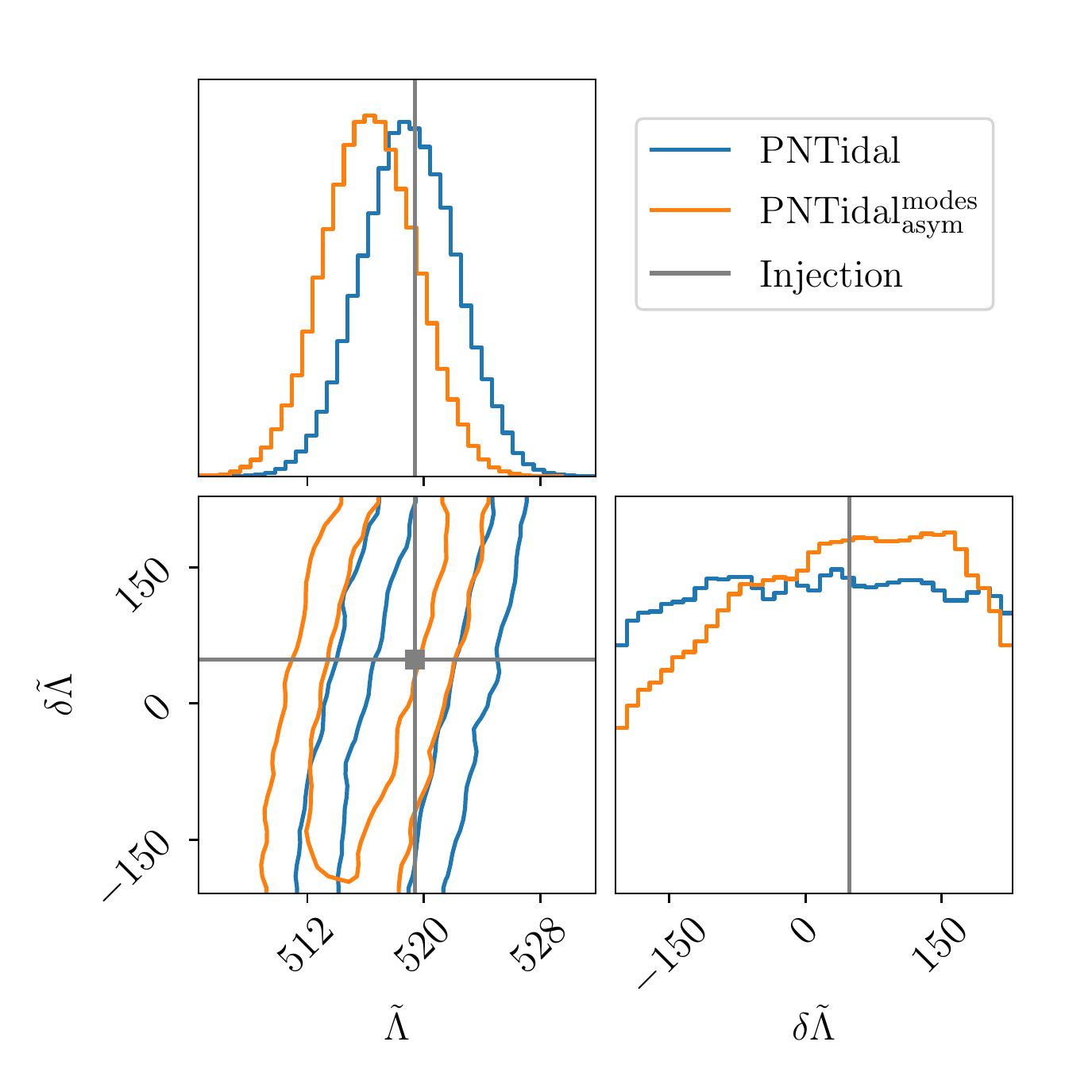}
    \caption{\emph{Gravitomagnetic effects for aligned spins of $\chi_{1,2}=0.005$} and SNR 1800. The blue curve corresponds to using the same waveform for injection and recovery. Comparing this with the orange curve indicates the changes due to gravitomagnetic tides from both the $m=1$ mode resonances and the adiabatic effects, which lead to a shift in the distribution of $\tilde \Lambda$ and a slight change in the shape of the $\delta \tilde \Lambda$ posterior. }
    \label{fig:recover_m1_spins}
\end{figure}
For small spins $\chi=0.005$, we see from Fig.~\ref{fig:recover_m1_spins} that the gravitomagnetic effects lead to a slightly larger shift in the posterior probability distributions than in the nonspinning case shown in Fig.~\ref{fig:recover_Pv2_nospin}. 
These trends become more discernible for higher spins of $\chi=0.01$ shown in Fig.~\ref{fig:recover_m1_spins_high}. For higher spins, the recovered distributions for $\tilde \Lambda$ with and without gravitomagnetic effects have essentially no overlap. We also notice that compared to the low-spin case in Fig.~\ref{fig:recover_m1_spins} the shift in the distribution for $\tilde \Lambda$ is in the opposite direction. We will investigate the causes of this below in Sec.~\ref{sec:discussion}. Roughly, it can be attributed to the fact that for higher spins the resonances occur at higher frequency, as seen in Fig.~\ref{fig:PSD}. Furthermore, as also found in~\cite{Ma:2020oni}, which included only the mode resonance effects with Newtonian parameters, the presence of gravitomagnetic tides significantly improves the measurability of $\delta \Lambda$. This is indicated by a peak in the one-dimensional projection or the size of the ellipse in the $\tilde\Lambda-\delta \tilde\Lambda$ plane, which is in contrast to the distribution being essentially uninformative when neglecting the gravitomagnetic effects (c.f. the blue curves in Fig~\ref{fig:recover_m1_spins_high}).

\begin{figure}[H]
\includegraphics[scale=0.635]{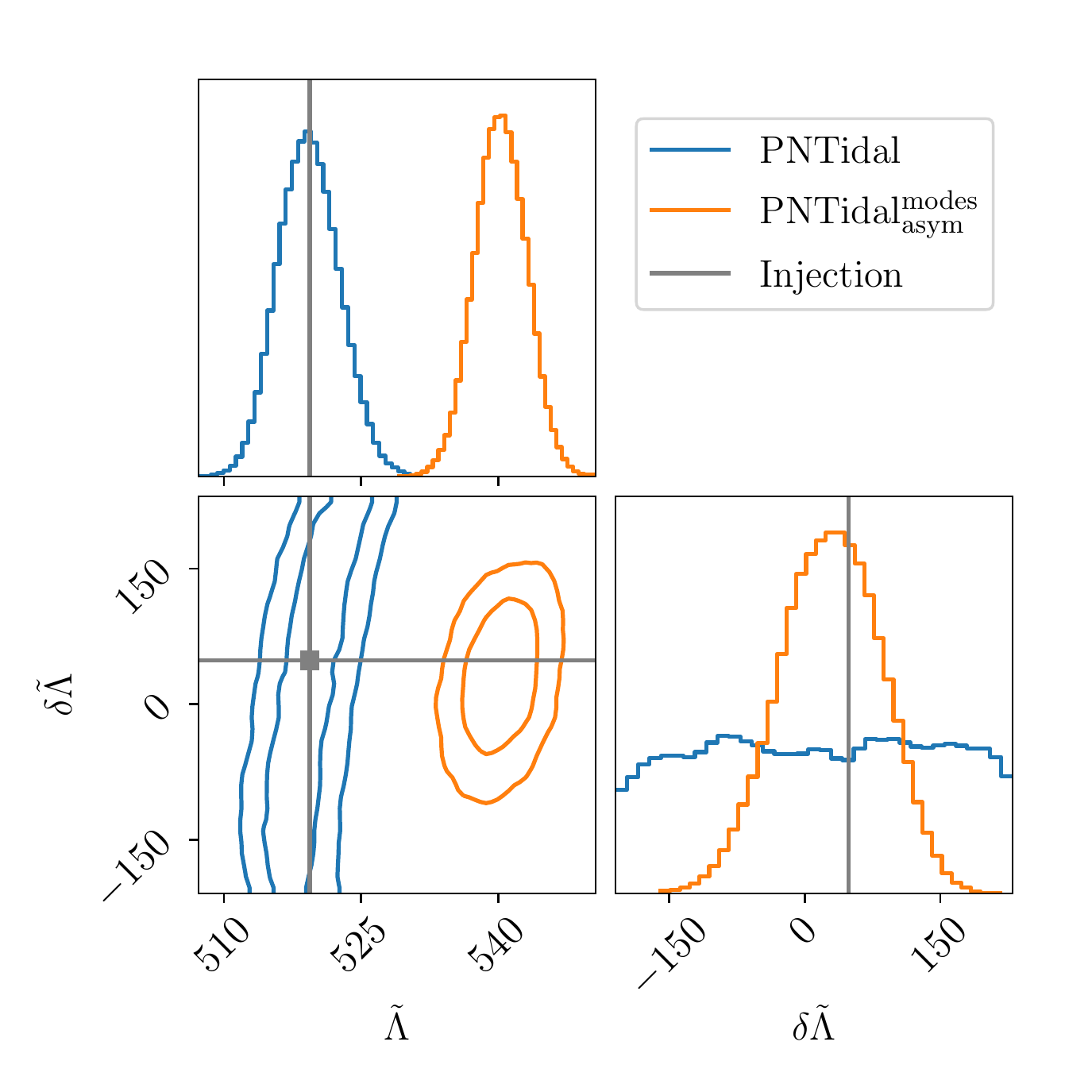}
 \caption{\emph{Gravitomagnetic effects for aligned spins of $\chi_{1,2}=0.01$} and SNR 1800. The blue curve corresponds to using the same waveform for injection and recovery, the orange curve indicates the effect of gravitomagnetic tides from both the $m=1$ mode resonances and the adiabatic effects. Significant shifts in $\tilde \Lambda$ and a peaked  shape of the distribution of $\delta \tilde \Lambda$ are clearly visible in this case. This is also illustrated by the two-dimensional representation of the error ellipses in the lower left panel. }
    \label{fig:recover_m1_spins_high}
\end{figure}

\subsubsection{Effects of different gravitomagnetic contributions for aligned spins}
\label{subsubsec:alignedspindissect}
\begin{figure}[H]
    \centering
    \includegraphics[scale=0.65]{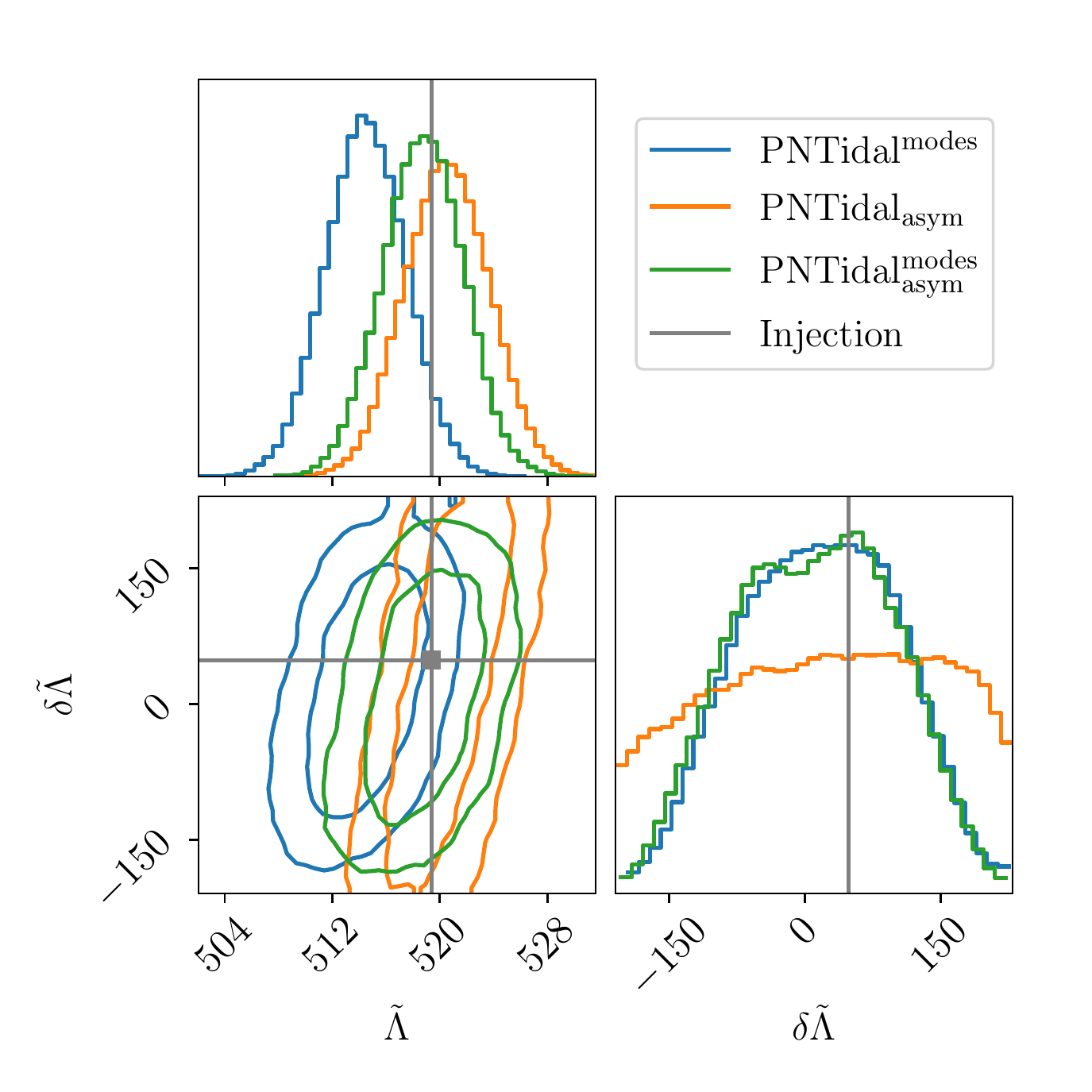}\\
    \includegraphics[scale=0.65]{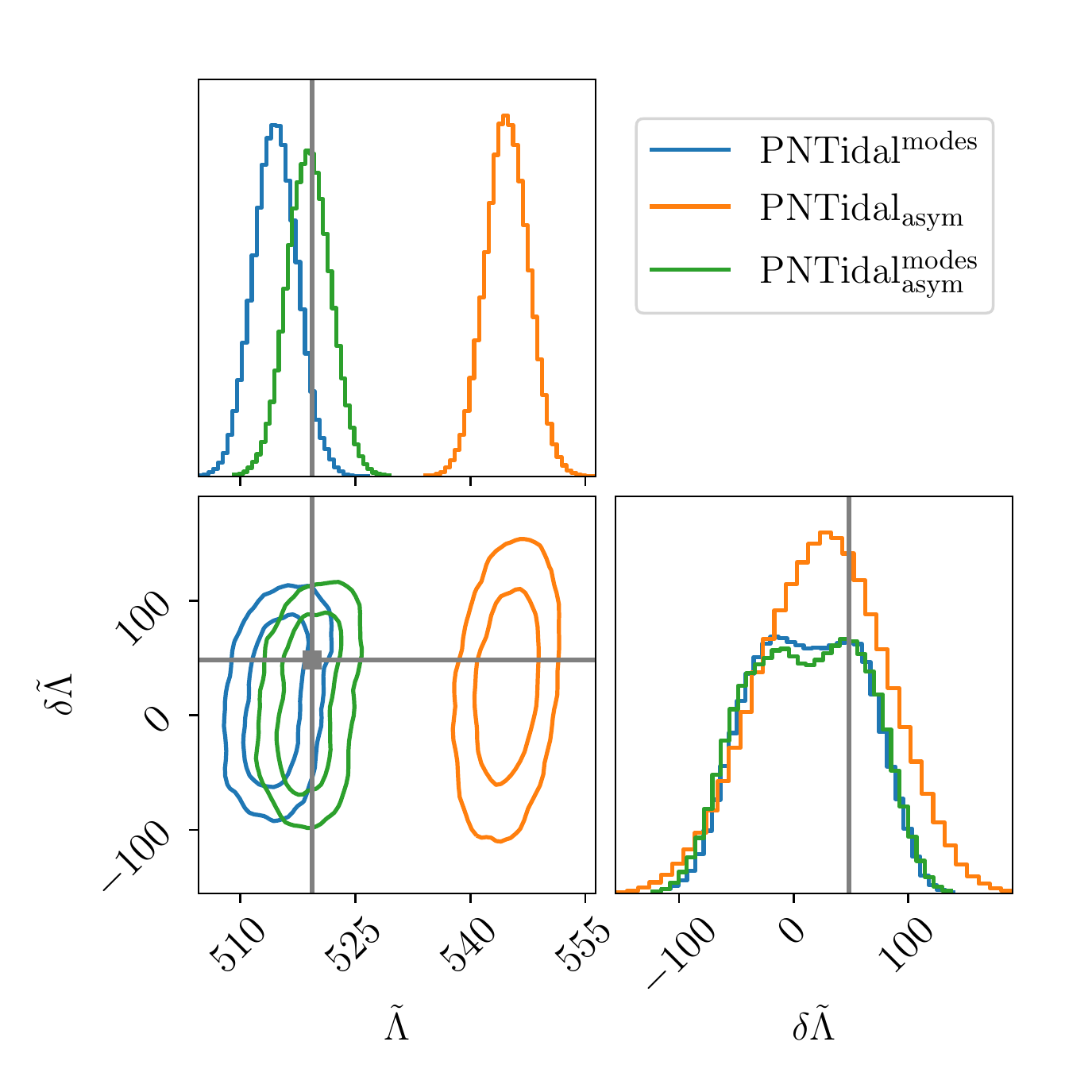}
    \caption{\emph{Effects of various gravitomagnetic contributions on the parameter recovery for aligned spins.} The results are for the systems with SNR 1800 and spins of  $\chi=0.005$ (upper panels) and  $\chi=0.01$ (lower panels). Green curves correspond to recovering with the same full model as used for the injection, blue curves include only the mode resonances, while orange curves indicate the adiabatic effects. We see that the conclusions about the impact of the resonance and adiabatic effect is opposite for the lower and higher spins: for low spins, adiabatic effects are most important for reducing the bias in $\tilde \Lambda$, while resonances give the dominant contribution to the measurability of $\delta \tilde \Lambda$. For high spins, the largest reduction in the bias in $\tilde \Lambda$ is due to the resonances, while the impact on $\delta \tilde \Lambda$ is comparable between resonance and adiabatic effects. }
    \label{fig:inject_m1_spins}
\end{figure}

Having quantified the impact of gravitomagnetic effects, we next investigate the relative importance of adiabatic and resonant contributions to these results. For this purpose, we switch to using the full tidal model PNTidal$_\text{asym}^\text{modes}$ for the injections. The results when recovering with different models that are missing various effects for the case with spins of $\chi=0.005$ are shown in the upper panels of Fig.~\ref{fig:inject_m1_spins}. The green curve illustrates the recovery with the same model as the injection, the blue curve corresponds to omitting the adiabatic effects, while the orange curve illustrates the omission of resonance effects from the model. From the large (small) shift away from the injected value in the distribution for $\tilde \Lambda$ when omitting (including) adiabatic effects it follows that the conclusions of Sec.~\ref{subsec:comparePani} about the signatures from adiabatic tides dominating over the resonance effects in this case continue to hold for the larger parameter space considered here. Furthermore, we also see by comparing the orange and blue curves in the upper panels of Fig.~\ref{fig:inject_m1_spins} that the more peaked distribution in $\delta \Lambda$ can be primarily attributed to the mode resonances in this case. 

The lower panel of Fig.~\ref{fig:inject_m1_spins} shows the same study with higher spins of $\chi=0.01$. We see the opposite behavior compared to the case with lower spins: now the mode resonances (blue curves) dominate over adiabatic effects (orange curve) for measuring $\tilde \Lambda$ without bias; in fact the inferred $\tilde \Lambda$ with the adiabatic model has no overlap with the injected value in this case. In all cases, a peaked  distribution in $\delta \Lambda $ emerges, indicating that it is measurable, though with significantly larger errors than $\tilde \Lambda$. The resonance effects yield a double-peaked distribution in this parameter for $\chi=0.01$, which we attribute to the larger spacing of the two resonances in this case. Interestingly, for $\chi=0.01$ the adiabatic effects contribute about equally to measuring $\delta\tilde \Lambda$ as the mode resonances, which is in contrast with the case of lower spins.

\subsubsection{Effect of the $m=2$ modes}
The analysis thus far focused on aligned-spin systems where only the $m=1$ modes are resonant. In this subsection we quantify the impact of the $m=2$ mode resonances by choosing spin orientations $\psi=\pi/3$ following a similar line of analysis as for the aligned-spin case. 
\begin{figure}[H]
    \centering
    \includegraphics[scale=0.635]{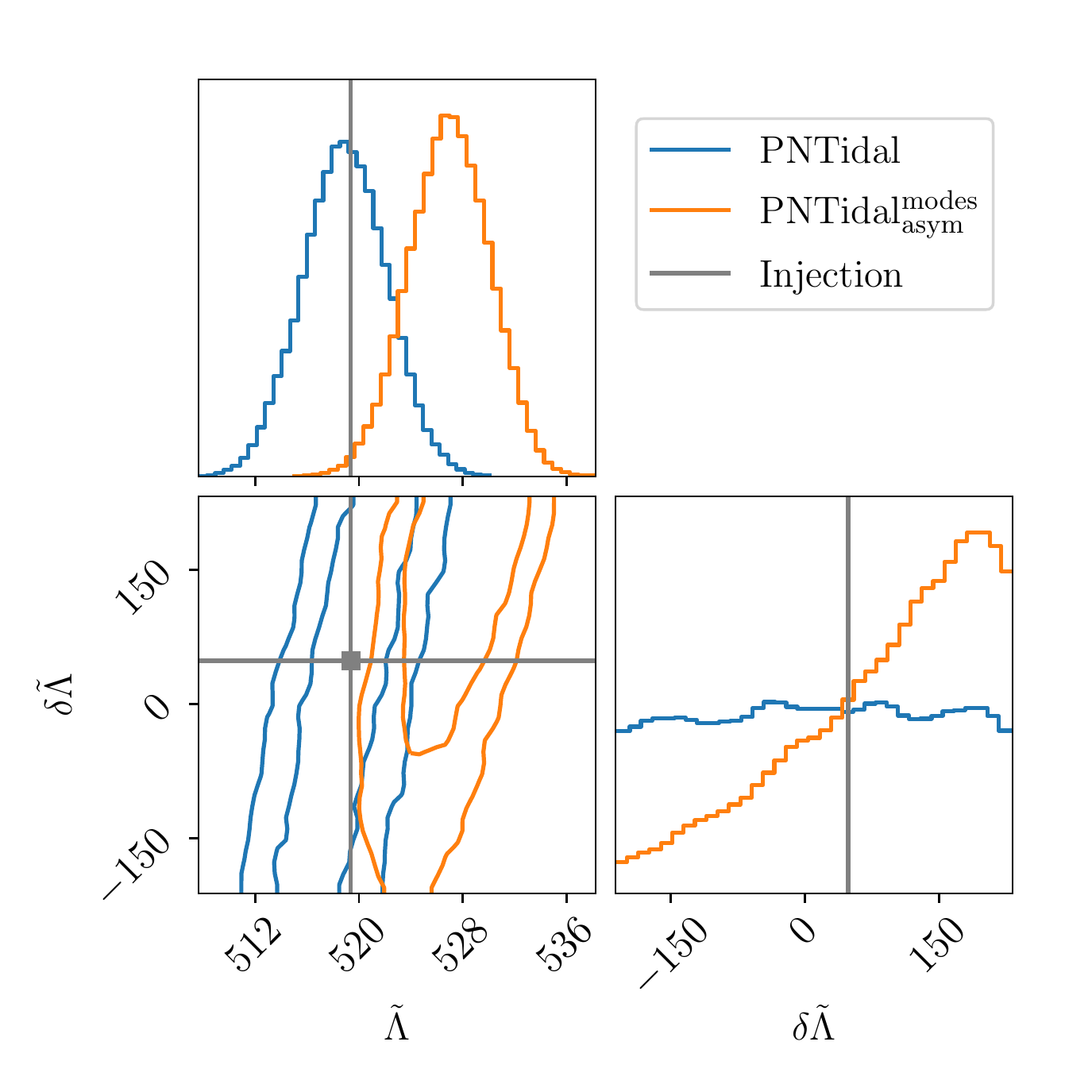}
       \caption{\emph{Gravitomagnetic effects for spin orientations $\psi=\pi/3$ and magnitudes $\chi=0.005$} at SNR 1800. The blue curve corresponds to using the same waveform for injection and recovery. Comparing this with the orange curve indicates the changes due to gravitomagnetic tides from both the $m=2$ mode resonances and the adiabatic effects, which lead to a shift in the distribution of $\tilde \Lambda$ and a more peaked shape of the $\delta \tilde \Lambda$ posterior.}
    \label{fig:recover_m2}
\end{figure}

First, we consider the impact of including all gravitomagnetic effects. From Fig.~\ref{fig:recover_m2} we see that even for small spins of $\chi=0.005$, the gravitomagnetic effects lead to larger shifts in the posterior probability distribution for $\tilde \Lambda$ and in the opposite direction compared to the aligned spin case in Fig.~\ref{fig:recover_m1_spins}. An approximate reasoning for this behavior is that the $m=2$ resonances occur later in the inspiral than the $m=1$ resonances, as we will discuss in more depth in Sec.~\ref{sec:discussion}.  
Figure~\ref{fig:recover_m2} also shows a peak in the distribution for $\delta\Lambda$ when including gravitomagnetic tides (orange curves), however, because the injection neglected gravitomagnetic effects, it is not centered on the injected value.

\begin{figure}[H]
    \centering
     \includegraphics[scale=0.635]{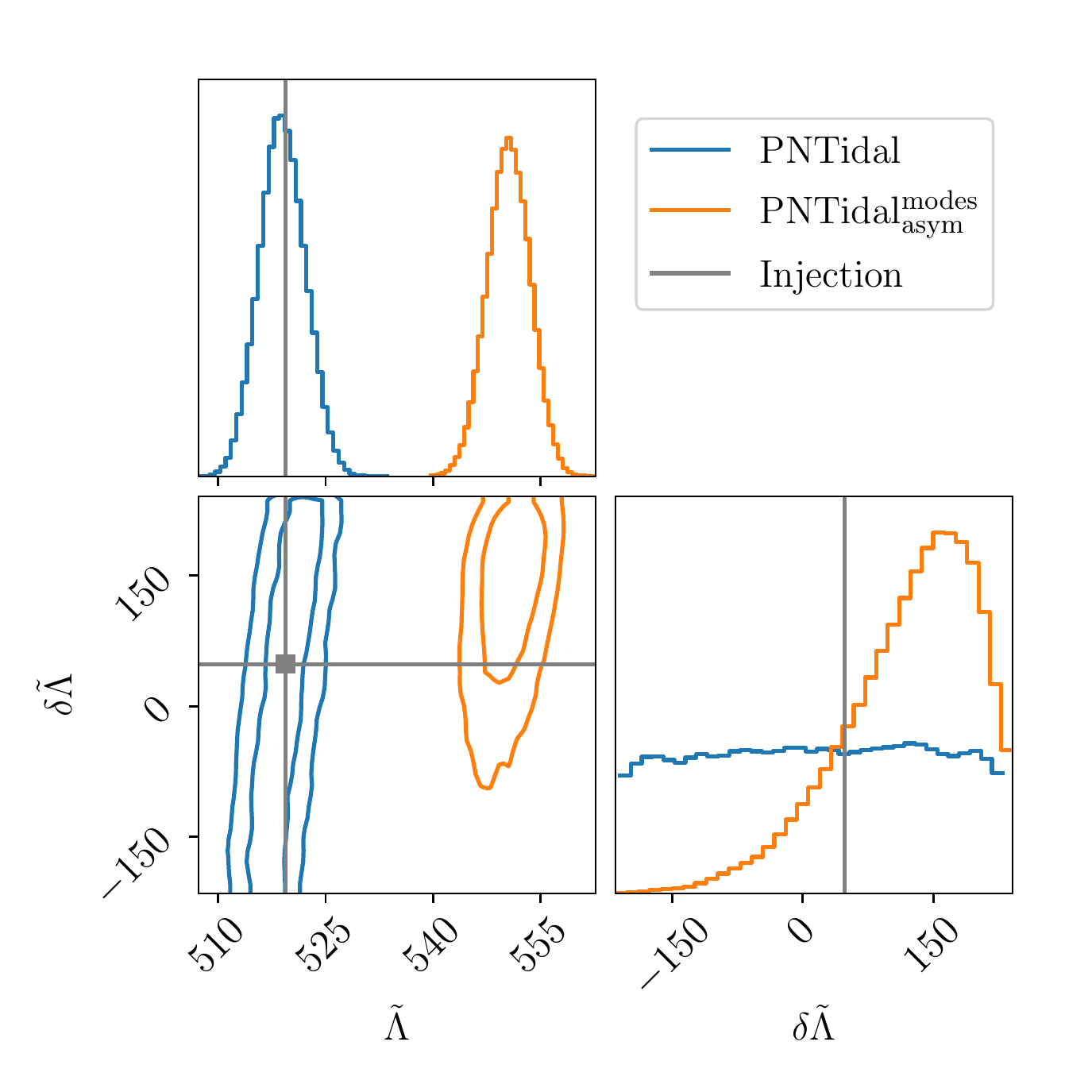}
    \caption{\emph{Gravitomagnetic effects for spin orientations $\psi=\pi/3$ and magnitudes $\chi_{1,2}=0.01$} at SNR 1800. The blue curve corresponds to using the same waveform for injection and recovery. Comparing this with the orange curve indicates the changes due to gravitomagnetic tides from both the $m=2$ mode resonances and the adiabatic effects, which lead to a substantial shift in the distribution of $\tilde \Lambda$ and clear peak in the $\delta \tilde \Lambda$ posterior.}
    \label{fig:recover_m2_high}
\end{figure}

For higher spins of $\chi=0.01$, the above trends are more pronounced, as seen in  Fig.~\ref{fig:recover_m2_high}. We observe that the two-dimensional confidence intervals have no overlaps at all in this case, and that the distribution in $\delta\Lambda$ becomes more distinctly peaked. 

\begin{figure}[H]
    \centering
    \includegraphics[scale=0.635]{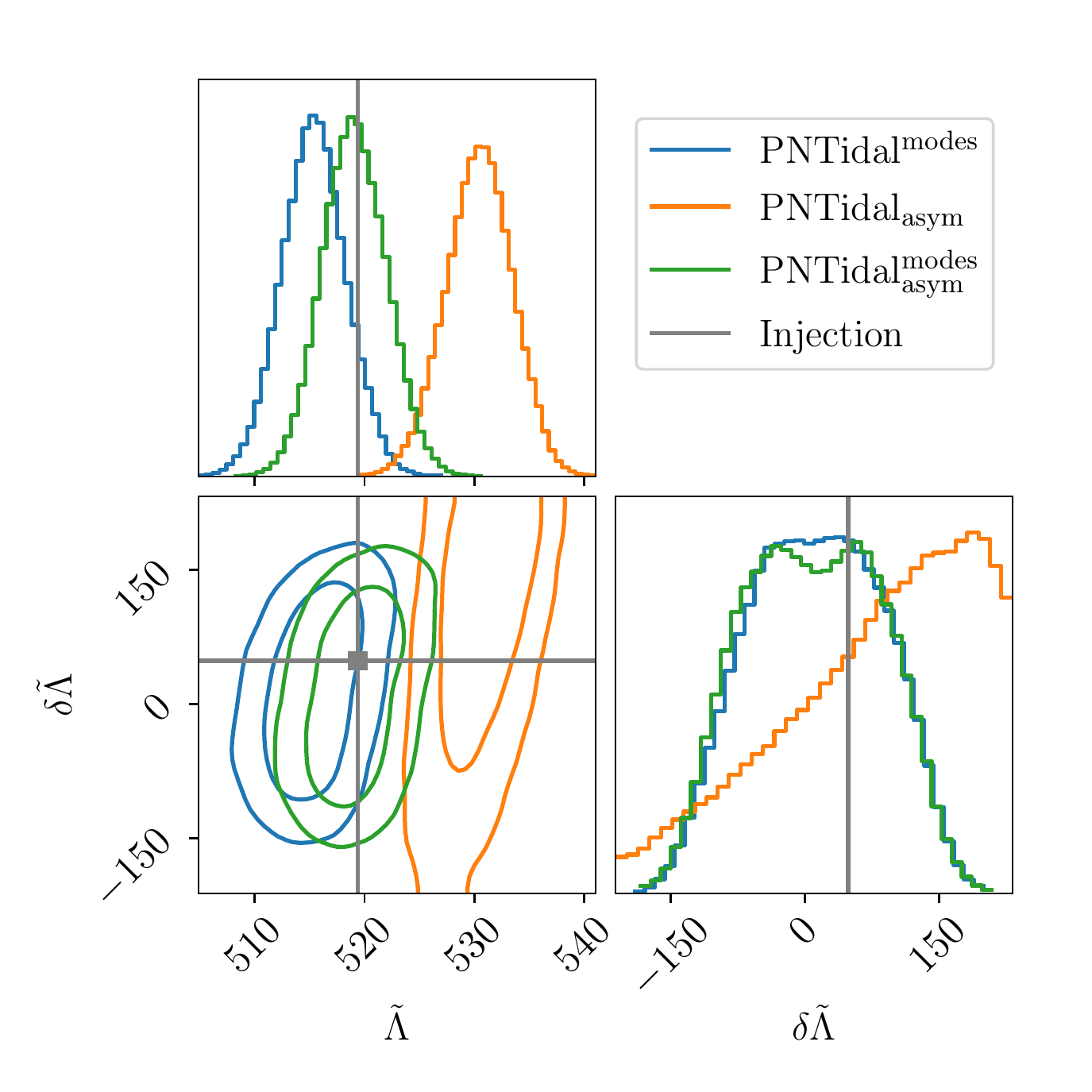}
    \includegraphics[scale=0.635]{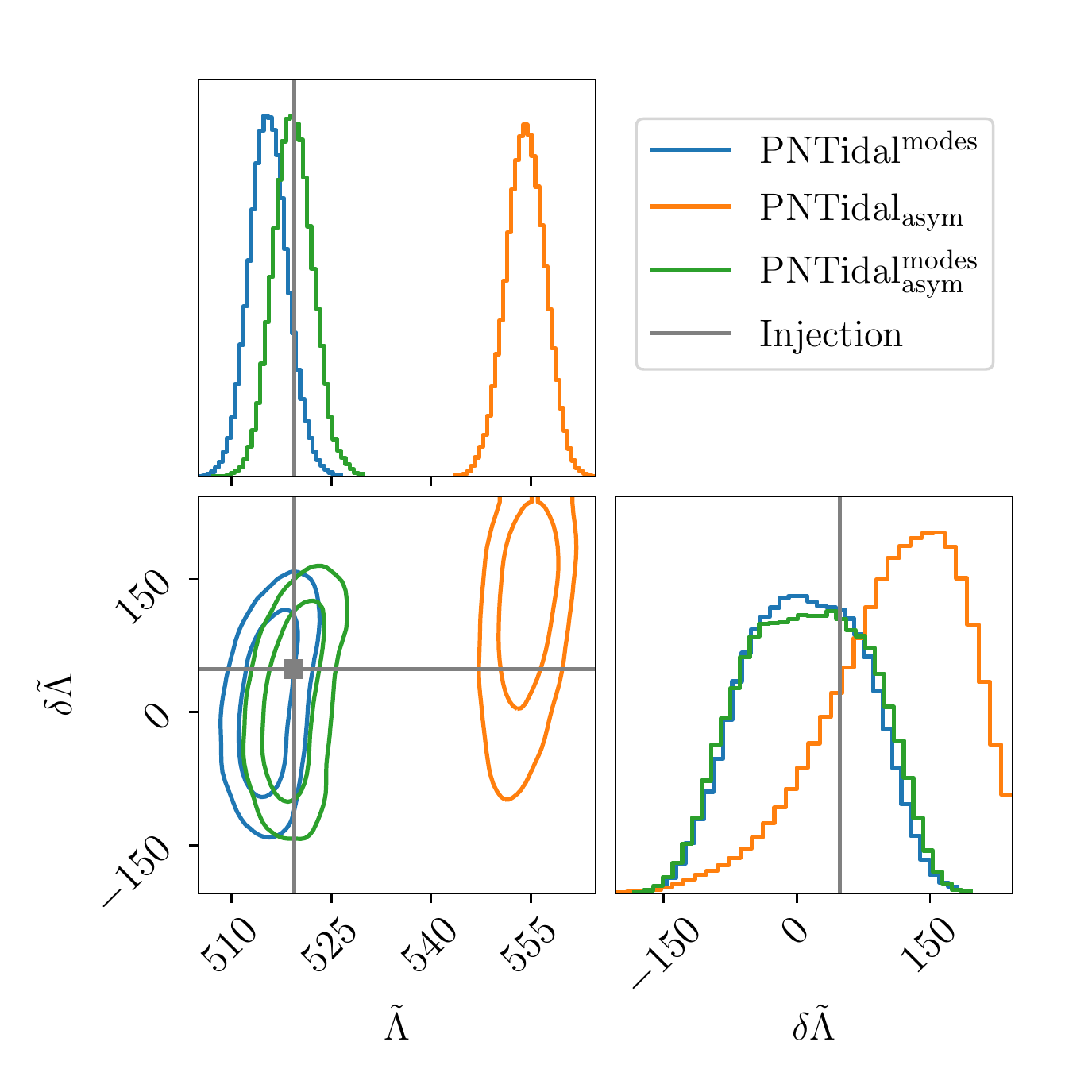}
    \caption{\emph{Effects of various gravitomagnetic contributions on the parameter recovery for misaligned spins.} The results are for the systems with SNR 1800 and spin orientations of $\psi=\pi/3$ with $\chi=0.005$ (upper panels) and  $\chi=0.01$ (lower panels). Green curves correspond to recovering with the same full model as the injection, blue curves include only the mode resonances, while orange curves indicate the adiabatic effects. We see that in both cases the mode resonances play a larger role for reducing biases than the adiabatic effects.}
    \label{fig:inject_m2_spin005}
\end{figure}
To gain deeper insights into the reasons for these results, we next characterize the impact of the resonant and adiabatic contributions to gravitomagnetic effects separately. The results of injecting with the full PNTidal$^{\rm asym}_{\rm modes}$ and recovering with different models for cases with smaller and larger spins are shown in the upper and lower panels of Fig.~\ref{fig:inject_m2_spin005} respectively. We see that the contributions of the $m=2$ mode resonances (blue curves) are more significant for reducing biases than the adiabatic effects (orange curves) for both the smaller and larger spin magnitudes in this case, though both effects are important to accurately recover the parameters.

\subsection{Measurement accuracy for different spins}
Having characterized the importance of the various contributions and gravitomagnetic tides overall, we next compare the net effects on the measurement accuracy for different spin magnitudes. 
\begin{figure}[H]
    \centering
    \includegraphics[scale=0.635]{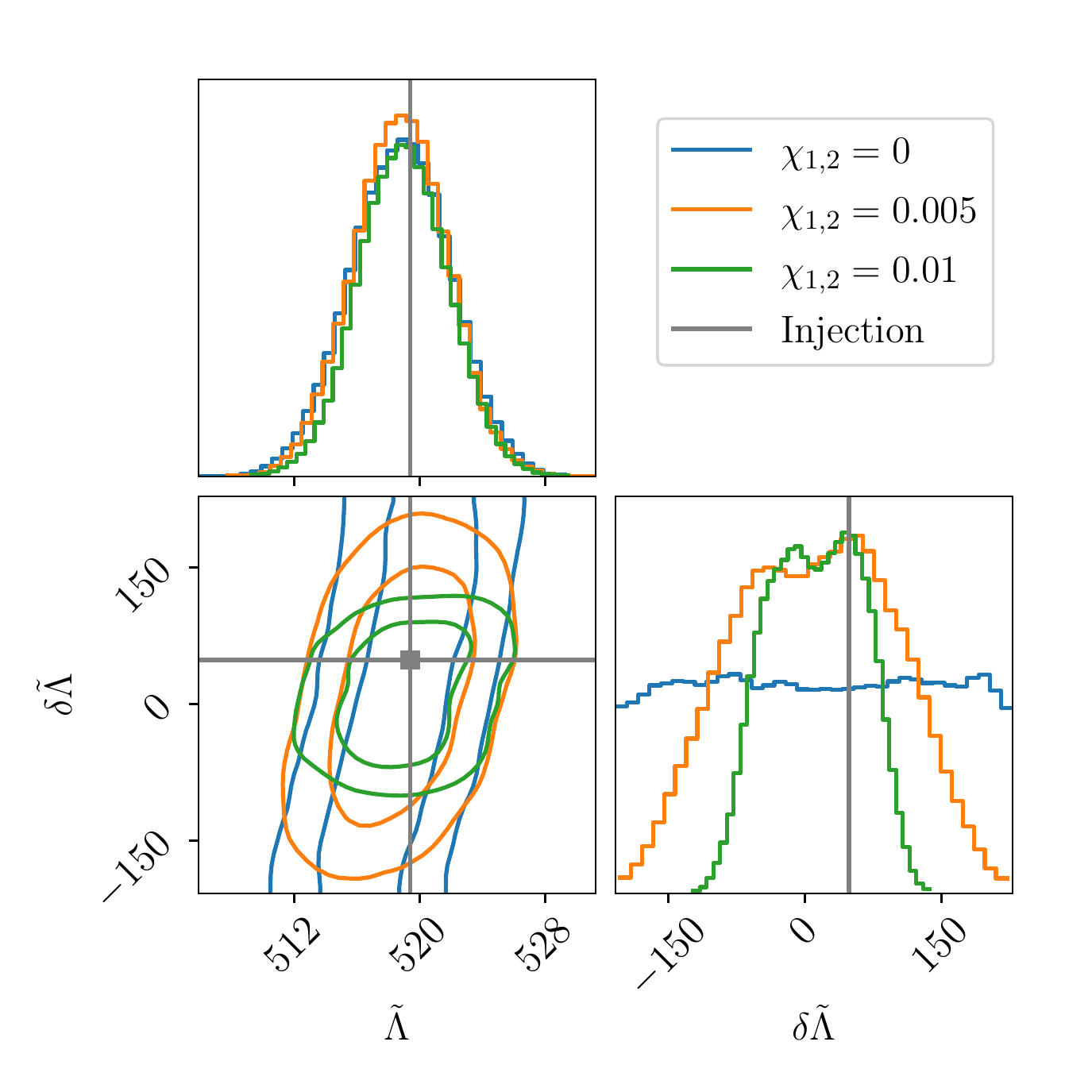}
    \caption{\emph{Effect of the spin magnitude on  inferred tidal parameters for aligned spins} and SNR of 1800. The  injection and recovery both use the same model PNTidal$_\text{asym}^\text{modes}$ with corresponding spin magnitudes, as indicated in the legend. Increasing the spin magnitude has very little impact on the width of the posterior in $\tilde \Lambda$ but significantly affects that of $\delta\tilde\Lambda$, where a higher spin leads to tighter bounds. }
    \label{fig:measurement1800_m1}
\end{figure}
In this study, the injected and recovered waveform is the full PNTidal$_\text{asym}^\text{modes}$ model with increasing spin $\chi=0, 0.005, 0.01$. The results for aligned spins are shown in Fig.~\ref{fig:measurement1800_m1} for increasing spin magnitudes $\chi=0, 0.005, 0.01$ corresponding to the blue, orange, and green curves respectively. We see that changing the spins has very little impact on the posterior distributions for $\tilde \Lambda$ in this case. By contrast, a decreasing spin results in a broader distribution in $\delta \tilde \Lambda$. As our analysis keeps the spins fixed, the impact of spins is through their coupling with adiabatic tidal parameters through~\eqref{eq:gravitoSigma}, the resonance phase shift $\sim \Omega^{2/3}$, and the mode resonance frequency, as we will further discuss in Sec.~\ref{sec:discussion}.
 
From the above results, we also infer that the double-peak in the distribution of $\delta \tilde \Lambda$ for $\chi=0.005$ arises from the combination of adiabatic and resonant effects, which act in opposite directions, while for $\chi=0.01$ it is largely due to the presence of two resonances spaced widely enough to be noticeable in the data analysis.   

\begin{figure*}
    \centering 
    \includegraphics[scale=0.97]{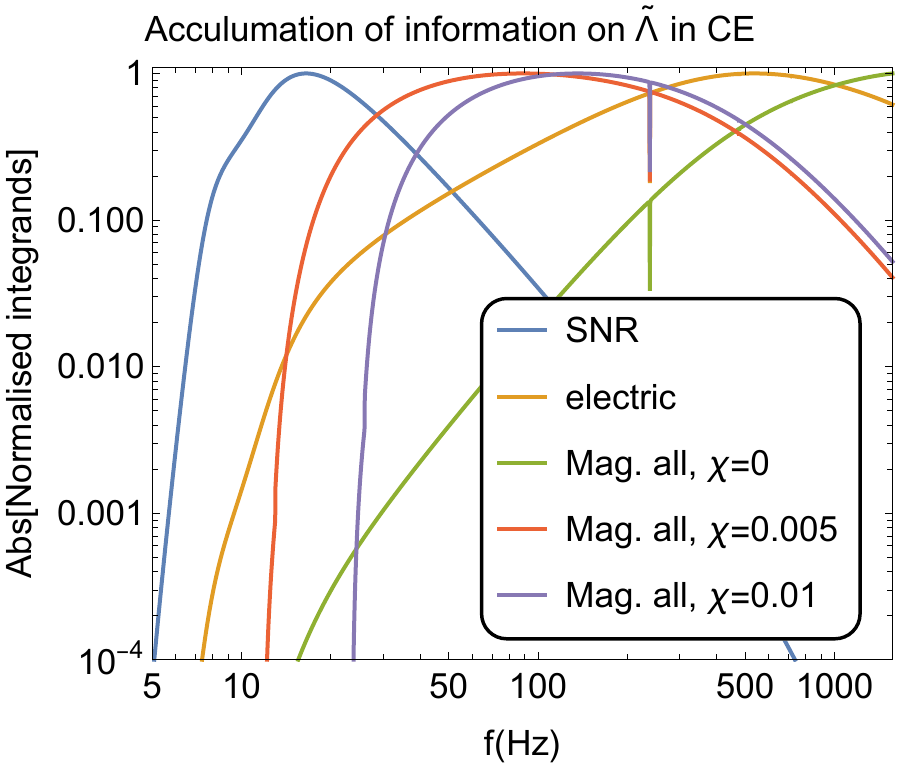}
    \includegraphics[scale=0.97]{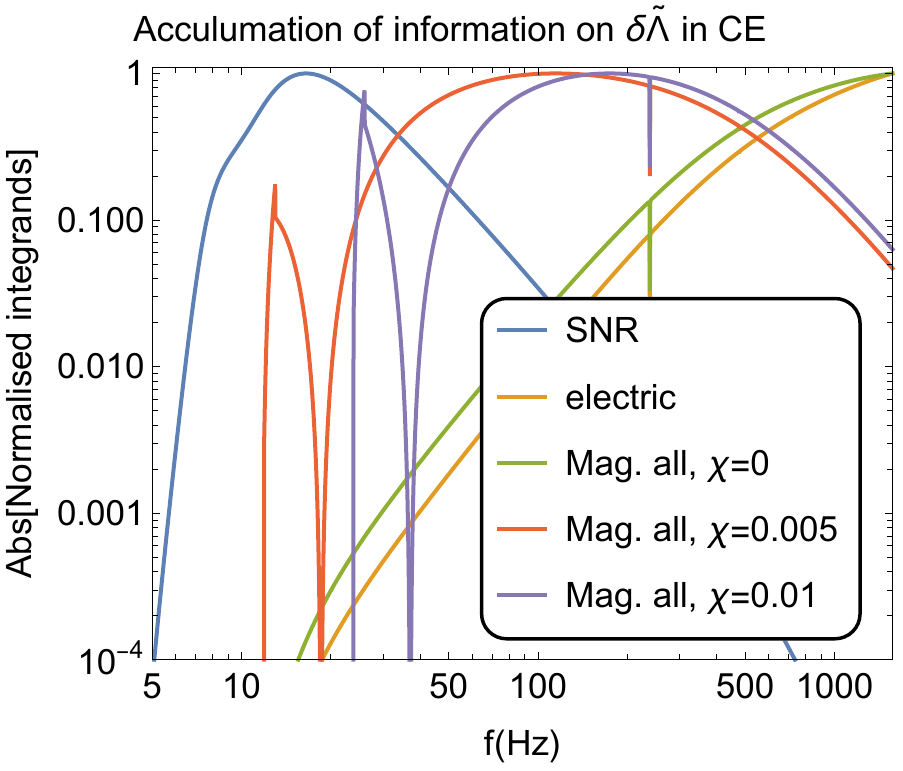}
    \caption{Accumulation of information encoded in integrands $\text{Abs}\left(\frac{\partial \tilde h^{\star}}{\partial \theta_i}\frac{\partial \tilde h}{\partial \theta_i}*\frac{1}{S_n(f)}\right)$ (normalized to its maximum value) for $\theta_i = \tilde\Lambda$ (left panel) and $\theta_i = \delta\tilde\Lambda$ (right panel) as a function of frequency for the injected value of aligned spins $\{0.0, 0.005, 0.01\}$, $\tilde\Lambda$= 519.38 and $\delta\tilde \Lambda = 48.37$. ``SNR'' denotes the integrands $\text{Abs}\left(\frac{\tilde h^{\star}\tilde h}{S_n(f)}\right)$, ``electric'' denotes only adiabatic gravitoelectric tidal contribution in~\eqref{eq:phasecontibutionfromtidal} and ``Mag. all" denotes adiabatic and resonant gravitomagnetic tidal contribution in~\eqref{eq:phasecontibutionfromtidal}.}
    \label{fig:accumulationofinformation}
\end{figure*}

A different perspective on the behavior can be gained by considering where in frequency the information about tidal parameters accumulates. This is not immediately visible from the phasing~\eqref{eq:phasecontibutionfromtidal} due to the implicit and nontrivial dependencies of the gravitomagnetic parameters on $\tilde \Lambda$ and $\delta \Lambda$ upon using the quasi-universal relations. Figure~\ref{fig:accumulationofinformation} shows the normalized integrands entering the Fisher matrix error computations. For $\tilde \Lambda$, the abrupt changes due to the resonances are too small to be visible on the scale of this plot, which is in contrast to the information on $\delta \tilde \Lambda$, where the resonance features are clearly visible. 

\begin{figure}[H]
    \centering
    \includegraphics[scale=0.635]{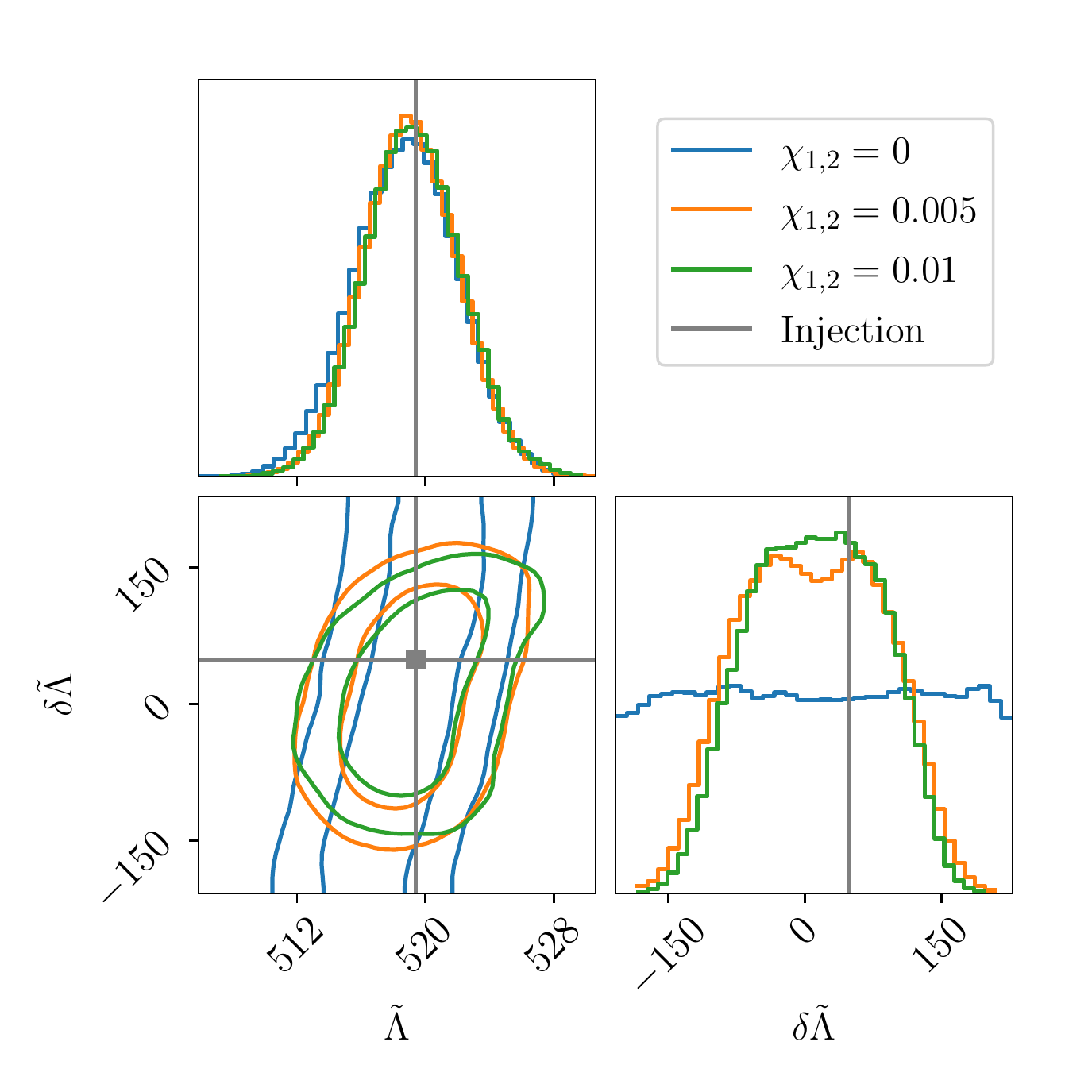}
    \caption{\emph{Effect of the spin magnitude on inferred tidal
parameters for inclined spins at $60^o$} and SNR of 1800. The injection and recovery both use the full model PNTidal$^{\rm modes}_{\rm
asym}$ with varying spin magnitudes as indicated in the legend. Increasing the spin magnitude from a finite value to a higher one has very little impact on the width of the posteriors in this case.}
    \label{fig:measurement1800_m2}
\end{figure}

The corresponding results with varying spin magnitudes for the case with misaligned spins of $\psi=\pi/3$ are shown in Fig.~\ref{fig:measurement1800_m2}.
We find similar trends as for the aligned spin case. However, a notable difference is that while the presence of spin has the expected impacts on the distributions, the consequences of any change in its magnitude are very small. This is in contrast with the trends in Fig.~\ref{fig:measurement1800_m1} for the $m=1$ modes. An explanation of this behavior could potentially come from considering the location of the resonances studied here with respect to the noise curve shown in Fig.~\ref{fig:PSD}, where changing the spin has a more drastic impact on the relative location of the $m=1$ resonances (diamonds) in the detector sensitivity. 

\begin{table}
\begin{tabular}{ |c|c|c|c |c}
\hline
 $\chi$ & $\psi=0$ & $\psi=\pi/3$ \\\hline
 \hline  & SNR 1800 &SNR 1800 \\\hline\hline
$0.0$  & $518.8_{-5.3}^{+5.0}$ $\left(10.5_{-194.2}^{+193.2}\right)$ & $518.8_{-5.3}^{+5.0}$ $\left(10.5_{-194.2}^{+193.2}\right)$ \\ 
\hline
$0.005$  & $518.8_{-4.9}^{+4.7}$ $\left(12.1_{-143.5}^{+135.5}\right)$ & $519.1_{-4.6}^{+4.8}$ $\left(9.2_{-118.4}^{+118.5}\right)$\\ 
\hline
$0.01$  & $519.0_{-4.5}^{+4.7}$ $\left(12.1_{-79.4}^{+75.6}\right)$
 & $519.2_{-4.6}^{+4.9}$ $\left(12.5_{-109.7}^{+106.8}\right)$\\
\hline\hline
& SNR 400 & SNR 400\\\hline
$0.0$  & $518.9_{-20.3}^{+20.6}$ ($14.6_{-198.8}^{+190.1}$) & $518.9^{+20.6}_{-20.3}$ ($14.6_{-198.8}^{+190.1}$)\\ 
\hline
$0.005$  & $519.1_{-20.5}^{+20.5}$ $\left(10.9_{-191.6}^{+191.9}\right)$ & $520.1_{-20.5}^{+20.1}$ $\left(12.2_{-191.8}^{+187.0}\right)$\\ 
\hline
$0.01$  & $520.4_{-20.0}^{+20.1}$ $\left(12.0_{-162.0}^{+158.5}\right)$
 & $521.1_{-19.4}^{+20.0}$ $\left(8.4_{-189.1}^{+188.3}\right)$\\\hline
\end{tabular}
\caption{ Recovered mean and 90\% credible intervals of $\tilde\Lambda ( \delta \tilde \Lambda )$ for SNR 1800 and 400. The injected values are $\tilde \Lambda=519$ and $\delta \Lambda =48$. The spin magnitude $\chi$ on each NS increases from top to bottom, and we recall that in the aligned spin case $\psi=0$ only the $m=1$ modes pass through resonance, for $\psi=\pi/3$ it is only the $m=2$ modes, and in the nonspinning case the resonances are absent. }\label{tab:snr1800errorbar}
\end{table}

\subsubsection{Extrapolating to lower SNR of 400}
\begin{figure*}
    \centering
    \includegraphics[scale=0.635]{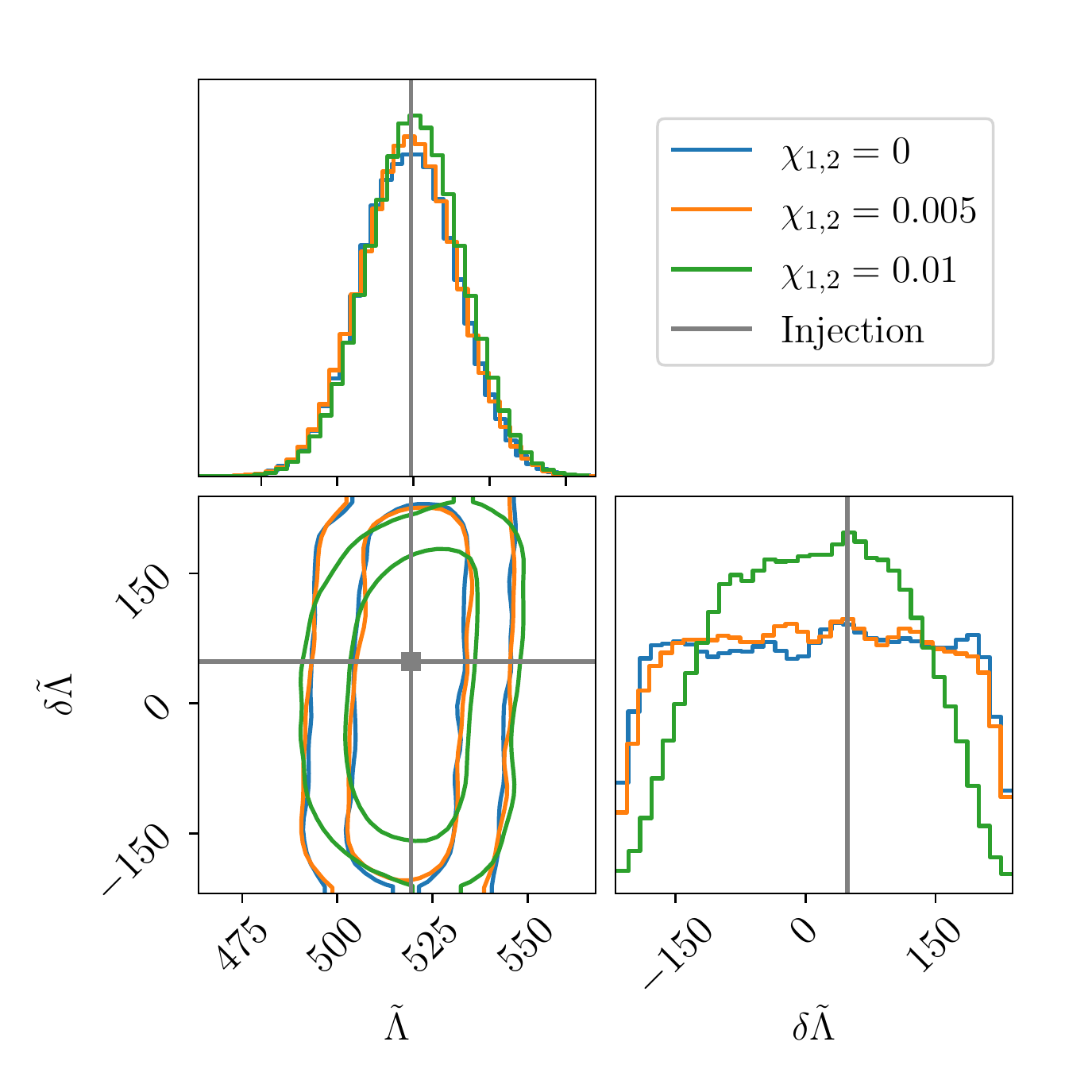}
    \includegraphics[scale=0.635]{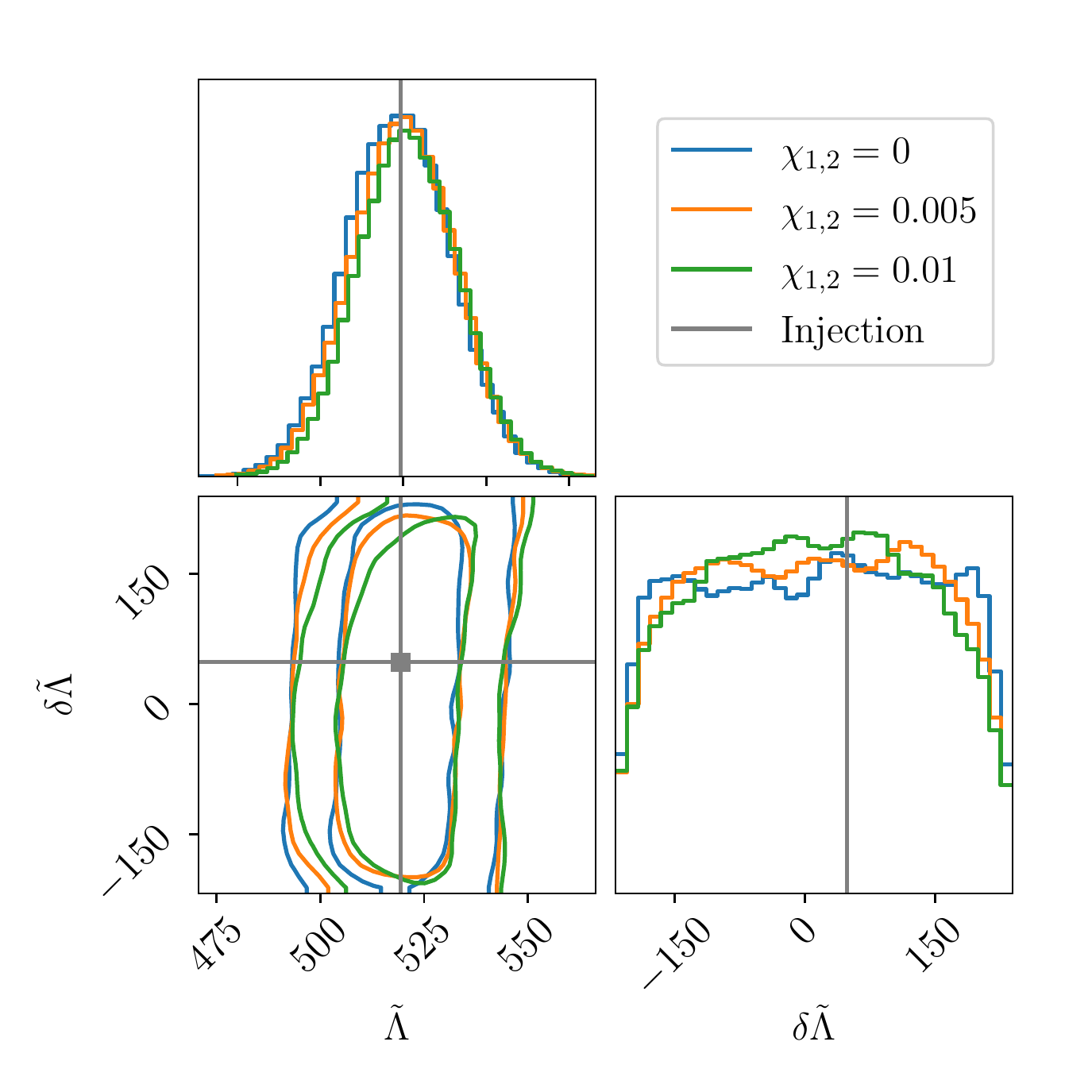}
    \caption{
    \emph{Fisher matrix results for systems with SNR 400 for different spins}.
    The injection and recovery both use the model PNTidal$_\text{asym}^\text{modes}$ with the corresponding spin magnitude indicated in the legend. \emph{Left panel}: aligned spins, \emph{right panel}: spin inclinations of $60^o$.
     Same as Figs.~\ref{fig:measurement1800_m1} and~\ref{fig:measurement1800_m2} except for lower SNR.
     }
 \label{fig:SNR400}
\end{figure*}

Thus far, we have assumed a SNR of 1800 in the CE detector, which is plausible for an event similar to GW170817. However, many more events will be observed at a lower SNR. To estimate the changes in our conclusions for such more numerous events, we perform the same analysis as above but for a SNR of 400 instead of 1800. From Fig.~\ref{fig:SNR400} we see that for lower SNR the qualitative trends of the effects of increasing the spins remain: there is little impact on the posterior distribution for $\tilde \Lambda$, while that for $\delta \tilde \Lambda $ becomes tighter. Comparing the left and right panels of Fig.~\ref{fig:SNR400}, which correspond respectively to the spin orientations where only the $m=1$ and $m=2$ modes are present, we also notice that for the higher spins considered here, the $m=1$ modes have a larger effect on the measurability of $\delta \Lambda$ than the $m=2$ modes. Comparing the results of Fig.~\ref{fig:SNR400} with the cases with higher SNR in Figs.~\ref{fig:measurement1800_m1} and~\ref{fig:measurement1800_m2} also quantifies the expected trends of a higher SNR resulting in tighter posterior distributions in the parameters. Table~\ref{tab:snr1800errorbar} lists the specific values obtained for the mean and $90\%$ credible intervals of the inferred $\tilde \Lambda$ and $\delta \tilde \Lambda$ distributions. From these results we see that for $\tilde \Lambda$, the change in the 90$\%$  interval for SNR 400 compared to 1800 is largely consistent with an approximate scaling of the errors as $({\rm SNR})^{-1}$, i.e. the width increases by roughly a factor of $\sim 4.5$. By contrast, the broadening of the $90\%$ interval in $\delta \tilde \Lambda$ with lower SNR is significantly less than expected from such a scaling, which is a promising indication for measurements, however, corroborating this for more realistic data analysis implications will require further work.

\section{Discussion}
\label{sec:discussion}

In this section, we discuss interesting aspects of the above findings and their interpretation. The high-level outcome of the case studies in Sec.~\ref{sec:results} is that they corroborate previously disconnected findings~\cite{Ma:2020oni, Jimenez-Forteza:2018buh} that gravitomagnetic tidal signatures in the GWs from both adiabatic and resonance-induced effects can have important impacts on the GW phasing for measurements with third-generation detectors. In addition, our analysis provided insights into the quantitative dependencies of these results on different features associated to the resonance-induced and adiabatic contributions and showed that their relative importance strongly depends on the system parameters. We discuss these findings below. 

\subsubsection{Features and parameter dependencies of gravitomagnetic effects in GWs}

\emph{Asymptotic adiabatic effects}. The leading-order contribution in the phase is parameterized by the quantity $\tilde \Sigma$ in~\eqref{eq:tidalphasing}, which increases slowly with $\tilde \Lambda$ and is positive both before and after a resonance. However, its magnitude significantly drops to much lower values across a resonance. In the GW phase, $\tilde \Sigma $ first enters together with $\delta \tilde\Lambda$ at the same scaling with frequency, and both with the opposite sign as the $\tilde \Lambda$ contribution, c.f.~\eqref{eq:tidalphasing}. These effects thus lead to a reduction of the net size of tidal GW signatures. Spin effects coupled with the adiabatic gravitomagnetic effects enter at a higher order in frequency through the parameter $\hat\Sigma$, thus contributing new information that breaks the degeneracy with $\delta \tilde \Lambda$. For the specific cases considered here, $\hat \Sigma$ is positive. The spin orientation impacts the size of the pre-resonance values of the adiabatic parameters $\tilde \Sigma $ and $\hat \Sigma$ 
, which can be larger for misaligned spins than for aligned spins.

\emph{Resonance-induced effects}. The resonance effects in the GW phase introduce a behavior that is very different from other contributions to the phasing because of its abruptness. Once present, the scaling with frequency is the same as for the gauge parameters $t_c, \phi_c$.
The size of the resonance-induced phase shifts depend on the spin magnitude and orientation, as well as the static and irrotational gravitomagnetic Love numbers characterizing how strongly the modes couple to the tidal field. The resonance jumps induce a negative GW phase correction, accelerating the inspiral and increasing the difference to a non-tidal signal. This is the opposite behavior as the leading-order adiabatic effects from gravitomagnetic tides discussed above. The resonance effects increase with larger $\tilde \Lambda$ and decrease for larger $\delta \tilde \Lambda$. Furthermore, larger spins lead to larger resonance jumps, as also seen from the spin dependence of~\eqref{eq:phasejump}, where $\Delta \Phi_{2m}\sim \chi^{2/3}$, and where we also note that the dependence on the spin orientation is such that $\Delta \Phi_{2m}$ is largest for aligned spins. In addition, the resonance frequencies are approximately proportional to the spin frequency as well as the mode number $m$. Larger spins and $m$ shift the resonances to higher frequencies, which can have several consequences depending on the resonance location. For example, for the case studies considered here, a shift of the resonances to higher frequencies leads to an enhanced accumulation of information from the pre-resonance adiabatic effects, the resonance jumps being within regimes of greater detector sensitivity, and a reduction in the number of cycles over which information from the resonances accumulates. As expected, when resonances occur within the most sensitive band of the detector, which in Sec.~\ref{sec:results} were the cases with $\chi=0.01$ and the scenario with $\chi=0.005$ with spins misaligned by $60^o$, the relative importance of the resonance effects is larger.

\subsubsection{Case studies of aligned-spin systems}
For systems with aligned spins, we found different trends depending on the spin magnitudes.
In the nonspinning case, only the adiabatic post-resonance effects contribute to the GW phase. As explained above, the leading-order adiabatic gravitomagnetic parameter $\tilde \Sigma$ contributes to the phasing~\eqref{eq:tidalphasing} in the same way as $\delta \tilde\Lambda$, while the contribution from $\hat\Sigma$ vanishes for zero spins. Consequently, gravitomagnetic effects have a rather small impact on the measurability of $\delta \tilde \Lambda$, as also seen in Fig.~\ref{fig:recover_Pv2_nospin}. Furthermore, the $\tilde \Sigma$-dependent contribution effectively reduces the size of the tidal effects in the phasing, which in this Fisher matrix study leads to the shift of the recovered $\tilde \Lambda$ to lower values, as also seen in Fig.~\ref{fig:recover_Pv2_nospin}.

For finite but low spins of $\chi=0.005$, the gravitomagnetic mode resonances occur at the lower end of CE's sensitive band, c.f. Fig~\ref{fig:PSD}, where the sensitivity is deteriorating. As seen in the upper panel of Fig.~\ref{fig:inject_m1_spins}, we find that in this case that the dominant contribution for recovering the correct mean for $\tilde \Lambda$ is the post-resonance asymptotic values. Consequently, the results for $\tilde \Lambda$ shown in
Fig.~\ref{fig:recover_m1_spins}, are similar to the nonspinning case in Fig.~\ref{fig:recover_Pv2_nospin}. When compared to the full baseline model with all gravitomagnetic effects, the mode resonances tend to lead to lower $\tilde \Lambda$ mean values, while adiabatic effects shift the distribution more towards higher ones in this case. A new feature with spins is that the $\delta \tilde \Lambda$ distribution becomes less flat, implying that this parameter becomes measurable, albeit with much larger statistical errors than $\tilde \Lambda$. As seen from  Fig.~\ref{fig:inject_m1_spins}, a non-flat distribution arises from both adiabatic effects and resonance jumps, however, the contribution from the latter is larger in this case.

For the higher spin system with $\chi=0.01$, where the resonances occur at higher frequencies, the posterior in $\tilde \Lambda$ with all gravitomagnetic effects is shifted in the opposite direction relative to the gravitoelectric baseline than the case with lower spins $\chi=0.005$, as seen by comparing Figs.~\ref{fig:recover_m1_spins} and~\ref{fig:recover_m1_spins_high}. This is due to the resonance effects becoming the dominant contribution to the results for $\tilde \Lambda$, as seen in the lower panel of Fig.~\ref{fig:inject_m1_spins}. 
Interestingly, the measurement of $\delta \tilde \Lambda$ in this higher-spin case is impacted nearly equally by both adiabatic and resonance effects, which both give similarly tight posteriors. We attribute the enhanced contribution from adiabatic effects for higher spins primarily to the larger contribution from $\hat \Sigma$, which breaks the degeneracies, with a potential further enhancement due to the resonance occuring at higher frequency, which increases the importance of the larger pre-resonance contribution to $\tilde \Sigma(\tilde \Lambda, \delta\tilde \Lambda)$, making the effects larger overall.

\subsubsection{Spins inclined at $60^o$}
The system with misaligned spins of $\chi=0.005$ we considered has the same resonance frequencies as the case study of aligned spins with $\chi=0.01$. However, the other parameters of these systems differ, which enables us to study their dependencies for a fixed resonance location. Specifically, the value of the phase jumps $\Delta \Phi_{2m}$ from~\eqref{eq:phasejump} are about four times larger for the high-spin $m=1$ case than for $m=2$ with low spin. Conversely, in the same comparison, $\tilde \Sigma$ is smaller by a factor of four and $|\hat\Sigma|$ is smaller by a factor of about two for the pre-resonance regime. The post-resonance values of $\tilde \Sigma $ are the same in the two cases. The outcomes of our analysis for the $m=2$ case with low spins are indeed qualitatively similar to that with the same resonance location but aligned spins. Notably, we find that the gravitomagnetic effects lead to a significant shift in $\tilde \Lambda$ compared to the gravitoelectric baseline and to a non-flat distribution of $\delta \tilde \Lambda$. Overall, the effects are larger for aligned high spins than for the misaligned low spin case, as can be seen by comparing, for example, Figs.~\ref{fig:recover_m1_spins_high} and~\ref{fig:recover_m2}.

For the case of misaligned spins with $\chi=0.01$, the results are similar to those with the lower spin magnitudes, as seen in Fig.~\ref{fig:measurement1800_m2}. This is in contrast with the aligned-spin case, where an increasing spin magnitude changes the importance of different gravitomagnetic contributions and noticeably improves the measurability as seen in Fig.~\ref{fig:measurement1800_m1}, for reasons explained above.

\section{Conclusion}
\label{sec:summary}
In this work, we  developed an approximate but efficient adaptation of known results to incorporate more realistic descriptions of resonant and adiabatic gravitomagnetic tidal effects in the Fourier-domain GW phasing for slowly rotating neutron stars and focusing on the quadrupole effects. We discussed the subtleties with adiabatic effects in this case, where calculations based on relativistic perturbation theory identified two different characteristic tidal deformability parameters. We derived the combinations of these parameters that appear together with a dependence on the spin orientation and the normalized mode frequencies in the GW signals and emphasized that they are different before and after a mode resonance. We also showed how to adapt an existing model for the resonance-induced GW phase shift to incorporate the fully relativistic properties of the neutron stars. In general, each neutron star passes through two quadrupolar gravitomagnetic resonances corresponding to the $m=1$ and $m=2$ modes, which for spins of $\chi \gtrsim 0.005$ lie within the sensitive band of third-generation GW detectors.

We used the above model to perform a data analysis study of the impact of gravitomagnetic effects on the measurements of tidal parameters with third-generation GW detectors, 
which relied on several simplifying assumptions. In particular, we used quasi-universal relations to reduce all matter parameters to the two tidal deformabilities, considered neutron stars with slightly unequal masses but equal spins and -orientations such that only one set of modes is resonantly excited during the inspiral, and mainly adopted a MCMC approach restricted to a four-dimensional subspace of parameters for GW170817-like events. These case studies enabled us to gain several quantitative insights and demonstrated that gravitomagnetic tides can be important to avoid biases in the inferred $\tilde \Lambda$ and lead to a peaked distribution in $\delta \tilde \Lambda$, which is flat and thus uninformative when neglecting gravitomagnetic effects.

To gain further insights into the underlying reasons for these results, we analyzed the different contributions to the gravitomagnetic tides, adiabatic versus resonance-induced, and compared the impacts from the $m=1$ modes (relevant for the aligned-spin configuration) to those of the $m=2$ modes (relevant for the case with misaligned spins). We found that for the $m=1$ modes, increasing the spin leads to increasingly better measurements of the tidal parameters. Furthermore, for aligned spins of magnitude $\chi=0.005$, the adiabatic effects are most important to avoid biases in the parameter $\tilde \Lambda$, while for larger spins of $\chi=0.01$ it is the mode resonances. In all cases, the mode resonances have a significant impact on the measurability of $\delta \tilde \Lambda$. On the other hand, for spin orientations where only the $m=2$ modes are resonant, we found no significant changes in the results with increasing spins. We also considered a case with a lower SNR of 400, as is expected for a larger number of events, and found  that similar qualitative trends persist. Interestingly, we noticed that while the broadening of the inferred posterior probability distribution seems to scale inversely with the SNR,  the broadening of the posterior in $\delta \tilde \Lambda$ is much less than this scaling. This is a promising indication for future measurements but will need to be confirmed with more realistic data analysis studies.

In conclusion, our work represents an exploratory study based on more realistic modeling of gravitomagnetic tides than in previous work. We made several simplifying assumptions and approximations, and neglected a number of additional matter effects that impact the GWs. Our results about the importance of the gravitomagnetic effects for measurements motivate more detailed data analysis studies as well as further advances in the modeling, which we leave to future work.  

\acknowledgements

This work was supported by the Netherlands Organization for Scientific Research (NWO). T.H. acknowledges support from the NWO sectorplan.

\appendix

\bibliography{main,main_auto}

\end{document}